\documentclass[aps,prd,twocolumn,amssymb,showpacs,superscriptaddress,nofootinbib,10pt,longbibliography]{revtex4-1}

\usepackage{amsfonts}
\usepackage{latexsym}
\usepackage{yfonts}
\usepackage{amsmath}
\usepackage{amssymb}
\usepackage{amsthm}
\usepackage{mathrsfs}
\usepackage{upgreek}
\usepackage{bm}
\usepackage{dcolumn}
\usepackage{epsfig}
\usepackage{graphicx}
\usepackage[latin1]{inputenc}
\usepackage{latexsym}
\usepackage{rotating}
\usepackage{hyperref}
\usepackage{color}
\usepackage{soul}

\begin{document}

\title{Entropy theorems in classical mechanics, general relativity, and the gravitational two-body problem}


\author{Marius Oltean}

\affiliation{\mbox{Observatoire des Sciences de l'Univers en r\'{e}gion Centre (OSUC), Universit\'{e} d'Orl\'{e}ans,}\\
\mbox{1A rue de la F\'{e}rollerie, 45071 Orl\'{e}ans, France}}

\affiliation{\mbox{P\^{o}le de Physique, Collegium Sciences et Techniques (CoST), Universit\'{e} d'Orl\'{e}ans,}\\
\mbox{Rue de Chartres, 45100  Orl\'{e}ans, France}}

\affiliation{Laboratoire de Physique et Chimie de l'Environnement et de l'Espace (LPC2E), Centre National de la Recherche Scientifique (CNRS), 3A Avenue de la Recherche Scientifique, 45071 Orl\'{e}ans, France}

\affiliation{\mbox{Departament de F\'isica, Facultat de Ci\`{e}ncies, Universitat Aut\`{o}noma de Barcelona,}\\
\mbox{Edifici C, 08193 Cerdanyola del Vall\`{e}s, Spain}}

\affiliation{\mbox{Institut de Ci\`encies de l'Espai (CSIC-IEEC),}\\
\mbox{Campus UAB, Carrer de Can Magrans s/n, 08193 Cerdanyola del Vall\`es, Spain}}

\author{Luca Bonetti}

\affiliation{\mbox{Observatoire des Sciences de l'Univers en r\'{e}gion Centre (OSUC), Universit\'{e} d'Orl\'{e}ans,}\\
\mbox{1A rue de la F\'{e}rollerie, 45071 Orl\'{e}ans, France}}

\affiliation{\mbox{P\^{o}le de Physique, Collegium Sciences et Techniques (CoST), Universit\'{e} d'Orl\'{e}ans,}\\
\mbox{Rue de Chartres, 45100  Orl\'{e}ans, France}}

\affiliation{Laboratoire de Physique et Chimie de l'Environnement et de l'Espace (LPC2E), Centre National de la Recherche Scientifique (CNRS), 3A Avenue de la Recherche Scientifique, 45071 Orl\'{e}ans, France}

\author{Alessandro D.A.M. Spallicci}

\affiliation{\mbox{Observatoire des Sciences de l'Univers en r\'{e}gion Centre (OSUC), Universit\'{e} d'Orl\'{e}ans,}\\
\mbox{1A rue de la F\'{e}rollerie, 45071 Orl\'{e}ans, France}}

\affiliation{\mbox{P\^{o}le de Physique, Collegium Sciences et Techniques (CoST), Universit\'{e} d'Orl\'{e}ans,}\\
\mbox{Rue de Chartres, 45100  Orl\'{e}ans, France}}

\affiliation{Laboratoire de Physique et Chimie de l'Environnement et de l'Espace (LPC2E), Centre National de la Recherche Scientifique (CNRS), 3A Avenue de la Recherche Scientifique, 45071 Orl\'{e}ans, France}

\author{Carlos F. Sopuerta}

\affiliation{\mbox{Institut de Ci\`encies de l'Espai (CSIC-IEEC),}\\
\mbox{Campus UAB, Carrer de Can Magrans s/n, 08193 Cerdanyola del Vall\`es, Spain}}

\date{\today}

\begin{abstract}
In classical Hamiltonian theories, entropy may be understood either as a statistical property of canonical systems, or as a mechanical property, that is, as a monotonic function of the phase space along trajectories. In classical mechanics, there are theorems which have been proposed for proving the non-existence of entropy in the latter sense. We explicate, clarify and extend the proofs of these theorems to some standard matter (scalar and electromagnetic) field theories in curved spacetime, and then we show why these proofs fail in general relativity; due to properties of the gravitational Hamiltonian and phase space measures, the second law of thermodynamics holds. As a concrete application, we focus on the consequences of these results for the gravitational two-body problem, and in particular, we prove the non-compactness of the phase space of perturbed Schwarzschild-Droste spacetimes. We thus identify the lack of recurring orbits in phase space as a distinct sign of dissipation and hence entropy production.
\end{abstract}


\maketitle



\section{\label{sec:Intro}Introduction}

The problem of reconciling the second law of thermodynamics\footnote{``It is the only physical theory of universal content concerning which I am convinced that, within the framework of applicability of its basic concepts, it will never be overthrown.'' \cite{einstein}} with classical (deterministic) Hamiltonian evolution is among the oldest in fundamental physics \cite{sklar_physics_1995,davies_physics_1977,brown_boltzmanns_2009}. In the context of classical mechanics (CM), this question motivated much of the development of statistical thermodynamics in the second half of the 19th century. In the context of general relativity (GR), thermodynamic ideas have occupied -- and, very likely, will continue to occupy -- a central role in our understanding of black holes and efforts to develop a theory of quantum gravity. Indeed, much work in recent years has been expended relating GR and thermodynamics \cite{rovelli_general_2012}, be it in the form of ``entropic gravity'' proposals \cite{verlinde_origin_2010,putten_entropic_2011,carroll_what_2016} (which derive the Einstein equation from entropy formulas), or gravity-thermodynamics correspondences \cite{freidel_gravitational_2013,freidel_non-equilibrium_2014} (wherein entropy production in GR is derived from conservation equations, in analogy with classical fluid dynamics). And yet, there is presently little consensus on the general meaning of ``the entropy of a gravitational system'', and still less on the question of why -- purely as a consequence of the dynamical (Hamiltonian) equations of motion -- such an entropy should (strictly) monotonically increase in time, i.e. obey the second law of thermodynamics.

However one wishes to approach the issue of defining it, gravitational entropy should in some sense emerge from suitably defined (micro-)states associated with the degrees of freedom \textsl{not} of any matter content in spacetime, but of the gravitational field itself -- which, in GR, means the spacetime geometry -- or statistical properties thereof. Of course, we know of restricted situations in GR where we not only have entropy definitions which make sense, but which also manifestly obey the second law -- that is, in black hole thermodynamics. In particular, the black hole entropy is identified (up to proportionality) with its area, and hence, we have that the total entropy increases when, say, two initially separated black holes merge -- a process resulting, indeed, as a direct consequence of standard evolution of the equations of motion. What is noteworthy about this is that black hole entropy is thus understood not as a statistical idea, but directly as a functional on the phase space of GR (comprising degrees of freedom which are subject to deterministic canonical evolution).

In CM, the question of the statistical nature of entropy dominated many of the early debates on the origin of the second law of thermodynamics during the development of the kinetic theory of gases \cite{sklar_physics_1995}. Initial hopes, especially by Boltzmann \cite{boltzmann_weitere_1872}, were that entropy could in fact be understood as a (strictly monotonic) function on classical phase space. However, many objections soon appeared which rendered this view problematic -- the two most famous being the reversibility argument of Loschmidt \cite{loschmidt} and the recurrence theorem of Poincar\'{e} \cite{hp1890am}.

The Loschmidt reversibility argument, in essence, hinges upon the time-reversal symmetry of the canonical equations of motion, and hence, the ostensibly equal expectation of evolution towards or away from equilibrium. Yet, arguably, this is something which may be circumvented via a sufficiently convincing proposition for identifying the directionality of (some sort of) arrow of time -- and in fact, recent work \cite{barbour_gravitational_2013,barbour_identification_2014} shows how this can actually be done in the Newtonian $N$-body problem, leading in this context to a clearly defined ``gravitational'' arrow of time. For related work in a cosmological context, see \cite{Sahni2015,Sahni2012}.

The Poincar\'{e} recurrence argument, on the other hand, relies on a proof that any canonical system in a bounded phase space will always return arbitrarily close to its initial state (and moreover it will do so an unbounded number of times) \cite{arnold_mathematical_1997,luis_barreira_poincare_2006}. As the only other assumption needed for this proof is Liouville's theorem (which asserts that, in any Hamiltonian theory, the probability measure for a system to be found in an infinitesimal phase space volume is time independent), the only way for it to be potentially countered is by positing an unbounded phase space for all systems -- which clearly is not the case for situations such as an ideal gas in a box.

Such objections impelled the creators of kinetic theory, Maxwell and Boltzmann in particular, to abandon the attempt to understand entropy -- in what we may accordingly call a mechanical sense -- as a phase space function, and instead to conceive of it as a statistical notion whose origin is epistemic ignorance, i.e. observational uncertainty
of the underlying (deterministic) dynamics. The famous H-Theorem of Boltzmann \cite{boltzmann_weitere_1872}, which was in fact initially put forth for the purposes of expounding the former, became reinterpreted and propounded in the light of the latter.

Of course, later such a statistical conception of entropy came to be understood in the context of quantum mechanics via the von Neumann entropy (defined in terms of the density matrix of a quantum system) and also in the context of information theory via the Shannon entropy (defined in terms of probabilities of a generic random variable). Indeed, the meaning of the word ``entropy'' is now often taken to reflect an observer's knowledge (or ignorance) about the microstates of a system.

Thus, the question of why the second law of thermodynamics should hold in a Hamiltonian system may be construed within two possible formulations -- on the one hand, a \textsl{mechanical}, and on the other, a \textsl{statistical} point of view. Respectively, we can state these as follows.

~

\noindent {\bf Problem I:}
Does there exist a function (or functional, if we are dealing with a field theory) on phase space which monotonically increases along the orbits of the Hamiltonian flow?

~

\noindent {\bf Problem II:}
Does there exist a function of time, defined in a suitable way in terms of a probability density on phase space, which always has a non-negative time derivative in a Hamiltonian system?

~

In CM, it is Problem II that has received the most attention since the end of the 19th century. In fact, there has been significant work in recent years by mathematicians \cite{villani_h-theorem_2008,yau_work_2010} aimed at placing the statistical formulation of the H-Theorem on more rigorous footing, and thus at proving more persuasively that, using appropriate assumptions, the answer to Problem II is in fact \textsl{yes}. In contrast, after the early Loschmidt reversibility and Poincar\'{e} recurrence arguments, Problem I has received some less well-known responses to the effect of demonstrating (even more convincingly) that the answer to it \textsl{under certain conditions} (to be carefully elaborated) is actually \textsl{no}. In this paper, we will concern ourselves with two such types of responses to Problem I: first, what we call the \textsl{perturbative} approach, also proposed by Poincar\'{e} \cite{poincare_sur_1889}; and second, what we call the \textsl{topological} approach, due to Olsen \cite{olsen_classical_1993} and related to the recurrence theorem. In the former, one tries to Taylor expand the time derivative of a phase space function, computed via the Poisson bracket, about a hypothetical equilibrium point in phase space, and one obtains contradictions with its strict positivity away from equilibrium. We revisit the original paper of Poincar\'{e}, clarify the assumptions of the argument, and carefully carry out the proof which is -- excepting a sketch which makes it seem more trivial than it actually turns out to be -- omitted therein. We furthermore extend this theorem to matter fields -- in particular, a scalar and electromagnetic field -- in curved spacetime. In the topological approach, on the other hand, one uses topological properties of the phase space itself to prove non-existence of monotonic functions. We review the proof of Olsen, and discuss its connections with the recurrence theorem and more recent periodicity theorems in Hamiltonian systems from symplectic geometry.

In GR, one may consider similar lines of reasoning as in CM to attempt to answer Problems I and II. Naively, one might expect the same answer to Problem II, namely \textsl{yes} -- however, as we will argue later in greater detail, there are nontrivial mathematical issues that need to be circumvented here even in formulating it. For Problem I, as discussed, one might confidently expect the answer to also be \textsl{yes}. Therefore, although we do not yet know how to define entropy in GR with complete generality, we \textsl{can} at least ask why the proofs that furnish a negative answer to Problem I in CM fail here, and perhaps thereby gain fruitful insight into the essential features we should expect of such a definition. 

Following the perturbative approach, we will show that a Taylor-expanded Poisson bracket does not contain terms which satisfy definite inequalities (as they do in CM). The reason, as we will see, is that the second functional derivatives of the gravitational Hamiltonian can (unlike in CM) be both positive and negative, and so its curvature in phase space cannot be used to constrain (functionals of) the orbits; no contradiction arises here with the second law of thermodynamics. 

Following the topological approach, there are two points of view which may explicate why the proofs in CM do not carry over to GR. Firstly, it is believed that, in general, the phase space of GR is non-compact \cite{schiffrin_measure_2012}. Of course, this assertion depends on the nature of the degrees of freedom thought to be available in the spacetime under consideration, but even in very simple situations (such as cosmological spacetimes), it has been shown explicitly that the total phase space measure diverges. Physically, what this non-compactness implies is the freedom of a gravitational system to explore phase space unboundedly, without having to return (again and again) to its initial state. This leads us to the second (related) point of view as to why the topological proofs in CM fail in GR: namely, the non-recurrence of phase space orbits. Aside from trivial situations, solutions to the canonical equations of GR are typically non-cyclic (i.e. they do not close in phase space) permitting the existence of functionals which may thus increase along the Hamiltonian flow. In fact, to counter the Poincar\'{e} recurrence theorem in CM, there even exists a ``no-return'' theorem in GR \cite{tipler_general_1979,tipler_general_1980,newman_compact_1986} for spacetimes which admit compact Cauchy surfaces and satisfy suitable energy and genericity conditions; it broadly states that the spacetime cannot return, even arbitrarily close, to a previously occupied state. One might nonetheless expect non-recurrence to be a completely general feature of all (nontrivial) gravitational systems, including spacetimes with non-compact Cauchy surfaces.

A setting of particular interest for this discussion is the gravitational two-body problem. With the recent detection \cite{abbott_et_al._observation_2016} (and ongoing efforts towards further observations \cite{blanchet_mass_2011}) of gravitational waves from two-body systems, the emission of which ought to be closely related to entropy production, a precise understanding and quantification of the latter is becoming more and more salient. In the CM two-body (i.e. Kepler) problem, the consideration of Problem I clearly explains the lack of entropy production due to phase space compactness (for a given finite range of initial conditions). In the Newtonian $N$-body problem, where (as we will elaborate) neither the perturbative nor the topological proofs are applicable, the answer to Problem I was actually shown to be \textsl{yes} in \cite{barbour_gravitational_2013,barbour_identification_2014}. In GR, the two-body problem may be considered in the context of perturbed Schwarzschild-Droste (SD) spacetimes\footnote{Commonly, this is referred to simply as the ``Schwarzschild metric''. Yet, it has long gone unrecognized that Johannes Droste, then a doctoral student of Lorentz, discovered this metric independently and announced it only four months after Schwarzschild \cite{dro16a,dro16b,sc16,ro02}, so for the sake of historical fairness, we here use the nomenclature ``Schwarzschild-Droste metric'' instead.} (as is relevant, for instance, in the context of extreme-mass-ratio inspirals). Here, the phase space volume (symplectic) form has been explicitly computed in \cite{jezierski_energy_1999}. We will use this in this paper to show that in such spacetimes, the phase space is non-compact; hence there are no contradictions with non-recurrence or entropy production.

We structure this paper as follows. In Section \ref{sec:Setup}, we establish some basic notation for describing general (constrained) Hamiltonian systems. In sections \ref{sec:CM} and \ref{sec:GR}, we address Problem I via the approaches described in this introduction in CM and GR, respectively. Then in Section \ref{sec:2BP}, we apply our discussion to the gravitational two-body problem, and finally in section \ref{sec:Conclusions}, we conclude.

\section{\label{sec:Setup}Setup}

We begin by establishing some basic notation for describing general canonical theories
which will be pertinent for our discussion. Technical details and definitions are relegated to appendix \ref{sec:A}, and comprehensive expositions can be found in \cite{bojowald_canonical_2011,hanson_constrained_1976}

Let $\mathscr{Q}$ denote the space of admissible configurations for
any given classical system (of particles and/or fields), whose dynamics can be determined from a given Lagrangian function $L$. From this, one may cast the theory in canonical form: its
degrees of freedom are then represented by a phase space $\mathscr{P}$ (which is the cotangent bundle of $\mathscr{Q}$), and its
dynamics are determined by a Hamiltonian function $H:\mathscr{P}\rightarrow\mathbb{R}$. In the case of field theories, the term phase space ``function'' should be understood as ``functional'' (of the fields).

The phase space $\mathscr{P}$ is, by construction, a symplectic
manifold. Let $\bm{\omega}$ denote the symplectic form on $\mathscr{P}$, and $\bm{\Omega}$ the volume form obtained therefrom. (The latter, when integrated over $\mathscr{P}$, gives its total volume or measure, $\mu(\mathscr{P})=\int_{\mathscr{P}}\bm{\Omega}$.) Furthermore,
let $\bm{X}_{F}$ be the Hamiltonian
vector field  of any phase space function $F:\mathscr{P}\rightarrow\mathbb{R}$. The time evolution of the system through $\mathscr{P}$
is represented by the integral curves of this vector field for the Hamiltonian function, $\bm{X}_{H}$. We denote the Hamiltonian flow (i.e. the flow generated by $\bm{X}_{H}$), for some time interval $\mathscr{T}\subseteq\mathbb{R}$, by $\Phi_{t}:\mathscr{P}\times\mathscr{T}\rightarrow\mathscr{P}$.

The situation becomes more subtle if the system under consideration
is constrained (as is the case, for example, with Maxwellian EM or
GR). Physically, the existence of constraints in a theory means that
not all points in $\mathscr{P}$ are dynamically accessible: in general,
not all initial conditions are permissible, and not all points in
$\mathscr{P}$ can be reached from permissible initial conditions. The consequence is that one can no longer use the phase space $\mathscr{P}$, but must instead work with a reduced phase space $\mathscr{S}$, the symplectic form of which we denote by $\bm{\omega}|_{\mathscr{S}}$.
The details of how this must be constructed are offered in Appendix~\ref{sec:A}. 

In what follows, what will be important is the reduced phase
space measure $\mu(\mathscr{S})$. It can be computed by integrating the volume form $\bm{\Omega}|_{\mathscr{S}}$ of $\mathscr{S}$ (determined by $\bm{\omega}|_{\mathscr{S}}$),
\begin{equation}
\mu\left(\mathscr{S}\right)=\int_{\mathscr{S}}\bm{\Omega}|_{\mathscr{S}}.\label{eq:phase_space_measure}
\end{equation}
We will see that the topological approach towards the validity of the second law of thermodynamics (described in the introduction) relies crucially on whether or not this quantity is divergent.

In situations where the implication is clear, we may drop the term ``reduced'' when making statements about the reduced phase space. Finally, using all this, we can now restate more precisely the above two problems on the second law of thermodynamics.

~

\noindent {\bf Problem I:}
Does there exist any $S:\mathscr{S}\rightarrow\mathbb{R}$ that monotonically increases along the orbits of $\Phi_{t}$?

~

\noindent {\bf Problem II:}
Does there exist any $S:\mathscr{T}\rightarrow\mathbb{R}$, defined in a suitable way in terms of a probability density $\rho:\mathscr{S}\times\mathscr{T}\rightarrow[0,1]$, satisfying ${\rm d}S/{\rm d}t\geq0$ in a Hamiltonian system? [Traditionally, the definition taken here for entropy is (a coarse-grained version of) $S\left(t\right)=-\int_{\mathscr{P}}\bm{\Omega}\,\rho\ln\rho$, or its appropriate reduction to $\mathscr{S}$ if there are constraints.]

\section{\label{sec:CM}Entropy theorems in classical mechanics}

\subsection{\label{sec:3.1}Setup}

Classical particle mechanics with $N$ degrees of freedom~\cite{arnold_mathematical_1997} can be formulated
as a Lagrangian theory with an $N$-dimensional configuration space
$\mathscr{Q}$. This means that we will have a canonical theory on
a $2N$-dimensional phase space $\mathscr{P}$. We can choose canonical
coordinates $\left(q_{1},...,q_{N}\right)$ with conjugate momenta
$\left(p_{1},...,p_{N}\right)$ such that the symplectic form on $\mathscr{P}$
is given by 
\begin{equation}
\bm{\omega}=\sum_{j=1}^{N}{\rm d}p_{j}\wedge{\rm d}q_{j}.\label{eq:CM_symplectic_form}
\end{equation}
Then, the volume form on $\mathscr{P}$ is simply the $N$-th
exterior power of the symplectic form, in particular $\bm{\Omega}=[(-1)^{N(N-1)/2}/N!]\bm{\omega}^{\wedge N}$, and $\bm{X}_{H}$ is here given in coordinates by 
\begin{equation}
\bm{X}_{H}=\sum_{j=1}^{N}\left(\frac{\partial H}{\partial p_{j}}\frac{\partial}{\partial q_{j}}-\frac{\partial H}{\partial q_{j}}\frac{\partial}{\partial p_{j}}\right).\label{eq:CM_time-evolution_vector}
\end{equation}
The action of $\bm{X}_{H}$ on any phase space function
$F:\mathscr{P}\rightarrow\mathbb{R}$, called the Poisson bracket,
gives its time derivative: 
\begin{equation}
\dot{F}=\frac{{\rm d}F}{{\rm d}t}=\bm{X}_{H}\left(F\right)=\left\{ F,H\right\} .\label{eq:CM_Poisson_bracket_definition}
\end{equation}
We obtain from this $\dot{q}_{j}=\{q_{j},H\}=\partial H/\partial p_{j}$
and $\dot{p}_{j}=\{p_{j},H\}=-\partial H/\partial q_{j}$, which are the canonical equations of motion. Moreover, we have that the symplectic form of $\mathscr{P}$, and hence its volume form, are preserved along $\Phi_{t}$; in other words, we have $\mathcal{L}_{\bm{X}_{H}}\bm{\omega}=0=\mathcal{L}_{\bm{X}_{H}}\bm{\Omega}$, which is known as Liouville's theorem.

We now turn to addressing Problem I in CM -- that is, the question of whether there exists a function
$S:\mathscr{P}\rightarrow\mathbb{R}$ that behaves like entropy in
a classical Hamiltonian system.
Possibly the most well-known answer given to this is the Poincar\'{e} recurrence theorem, the proof of which we present in appendix \ref{sec:A}.

Let us now discuss, in turn, the perturbative and topological approaches.

\subsection{\label{sec:3.2}Perturbative approach}

We revisit and carefully explicate, in this subsection, the argument given by Poincar\'{e}~\cite{poincare_sur_1889} to the effect that an entropy function $S:\mathscr{P}\rightarrow\mathbb{R}$ does not exist.
First, we will clarify the assumptions that need to go into it, i.e. the conditions we must impose both on the entropy $S$ as well as on the Hamiltonian $H$, and then we will supply a rigorous proof.

\subsubsection{\label{sec:3.2.1}Review of Poincar\'{e}'s idea for a proof}

In his original paper \cite{poincare_sur_1889} (translated into English in~\cite{olsen_classical_1993}), the argument given by Poincar\'{e} (expressed using the contemporary notation of this paper) for the non-existence
of such a function $S:\mathscr{P}\rightarrow\mathbb{R}$ is the following:
if $S$ behaves indeed like entropy, it should satisfy 
\begin{equation}
\dot{S}=\left\{ S,H\right\} =\sum_{k=1}^{N}\left(\frac{\partial H}{\partial p_{k}}\frac{\partial S}{\partial q_{k}}-\frac{\partial H}{\partial q_{k}}\frac{\partial S}{\partial p_{k}}\right)>0\label{eq:CM_Sdot_Poisson_Poincare}
\end{equation}
around a hypothetical equilibrium point in $\mathscr{P}$. Taylor
expanding each term and assuming all first partials of $S$ and
$H$ vanish at this equilibrium, we obtain a quadratic form (in the
distances away from equilibrium) plus higher-order terms. If we are
``sufficiently close'' to equilibrium, we may ignore the higher-order
terms and simply consider the quadratic form, which thus needs to
be positive definite for the above inequality [Eq.~\eqref{eq:CM_Sdot_Poisson_Poincare}]
to hold. But here Poincar\'{e}, without presenting any further explicit
computations, simply asserts that ``it is easy to satisfy oneself
that this is impossible if one or the other of the two \textit{quadratic forms $S$
and $H$} is definite, which is the case here.'' (Our modern language modification is in italic.)

Neither the casual dismissal of the higher-order terms, nor, even
more crucially, the fact that ``it is easy to satisfy oneself''
of the impossibility of this quadratic form to be positive definite
is immediately apparent from this discussion. In fact, all of the
points in this line of reasoning require a careful statement of the
necessary assumptions as well as some rather non-trivial details of
the argumentation required to obtain the conclusion (that $\dot{S}=0$).

In what follows, we undertake precisely that. First we look at the
assumptions needed for this method to yield a useful proof, and then
we carry out the proof in full detail and rigor.

\subsubsection{\label{sec:3.2.2}Entropy conditions}

A function $S:\mathscr{P}\rightarrow\mathbb{R}$ can be said to behave
like entropy insofar as it satisfies the laws of thermodynamics. In
particular, it should conform to two assumptions: first, that it should
have an equilibrium point, and second, that it should obey the second
law of thermodynamics -- which heuristically states that it should be
increasing in time everywhere except at the equilibrium point, where
it should cease to change. We state these explicitly as follows: 

~

\noindent{\bf S1} \textsl{(Existence of equilibrium)}{\bf:} We assume there exists a point
in phase space, $x_{0}\in\mathscr{P}$, henceforth referred to as
the ``equilibrium'' configuration of the system, which is a stationary
point of the entropy $S$, i.e. all first partials thereof should
vanish when evaluated there: 
\begin{equation}
\left(\frac{\partial S}{\partial q_{j}}\right)_{0}=0=\left(\frac{\partial S}{\partial p_{j}}\right)_{0},\label{eq:CM_Equilibrium}
\end{equation}
where, for convenience, we use the notation $\left(\cdot\right)_{0}=\left.\left(\cdot\right)\right|_{x_{0}}$
to indicate quantities evaluated at equilibrium. Note that by the
definition of the Poisson bracket [Eq.~\eqref{eq:CM_Poisson_bracket_definition}],
this implies $(\dot{S})_{0}=0$. 

~

\noindent{\bf S2} \textsl{(Second law of thermodynamics)}{\bf:} A common formulation of the
second law asserts that the entropy $S$ is always increasing in time
when the system is away from equilibrium (i.e. $\dot{S}>0$ everywhere
in $\mathscr{P}\backslash x_{0}$), and attains its maximum value
at equilibrium, where it ceases to change in time (i.e. $\dot{S}=0$
at $x_{0}$, as implied by the first condition). We need to work,
however, with a slightly stronger version of the second law: namely,
the requirement that the Hessian matrix of $\dot{S}$, 
\begin{equation}
\mathbf{Hess}(\dot{S})=
\left[
\begin{array}{c|c}
\underset{}{{\displaystyle \frac{\partial^{2}\dot{S}}{\partial q_{i}\partial q_{j}}}} & \underset{}{{\displaystyle \frac{\partial^{2}\dot{S}}{\partial q_{i}\partial p_{j}}}} \\
\hline
\overset{}{{\displaystyle \frac{\partial^{2}\dot{S}}{\partial p_{i}\partial q_{j}}}} & \overset{}{{\displaystyle \frac{\partial^{2}\dot{S}}{\partial p_{i}\partial p_{j}}}}
\end{array}
\right],\label{eq:CM_Hessian_S_dot}
\end{equation}
is positive definite when evaluated at equilibrium, i.e. $(\mathbf{Hess}(\dot{S}))_{0}\succ0$. 

~

We make now a few remarks about these assumptions. Firstly, S2 is a sufficient -- though not strictly necessary -- condition to
guarantee $\dot{S}>0$ in $\mathscr{P}\backslash x_{0}$ and $\dot{S}=0$
at $x_{0}$. However, the assumption of positive definiteness of the Hessian
of the entropy $S$ itself at equilibrium is often used in statistical
mechanics \cite{abad_principles_2012}, and so it may not be objectionable
to extend this supposition to $\dot{S}$ as well. (In any case, this
leaves out only special situations where higher-order derivative tests
are needed to certify the global minimisation of $\dot{S}$ at equilibrium,
which arguably are more of mathematical rather than physical interest;
we may reasonably expect the entropy as well as its time derivative
to be quadratic in the phase space variables as a consequence of its
ordinary statistical mechanics definitions in terms of energy.)

Secondly, the above two conditions omit the consideration of functions on $\mathscr{P}$
which are \textsl{everywhere} strictly monotonically increasing in
time, i.e. whose time derivative is always positive with no equilibrium
point. The topological approaches to Problem I, which we will turn to in the next subsection, do accommodate the possibility such
functions.

Thirdly, the equilibrium point $x_{0}\in\mathscr{P}$, though usually (physically)
expected to be unique, need not be for the purposes of what follows,
so long as it obeys the two conditions S1 and S2. In other words,
it suffices that there exists at least one such point in $\mathscr{P}$.

Fourthly, there is no topological requirement being imposed on
the phase space $\mathscr{P}$. It is possible, in other words, for
its total measure $\mu\left(\mathscr{P}\right)=\int_{\mathscr{P}}\bm{\Omega}$
to diverge. This means that the theorem applies to systems which can, in principle, explore phase space unboundedly, without any limits being imposed (either physically or mathematically) thereon.

\subsubsection{\label{sec:3.2.3}Hamiltonian conditions}

Next, we make a few assumptions about the Hamiltonian $H:\mathscr{P}\rightarrow\mathbb{R}$
which we need to impose in order to carry out our proof. The first
two assumptions are reasonable for any typical Hamiltonian in classical
mechanics, as we will discuss. The third, however, is stronger than
necessary to account for all Hamiltonians in general -- and indeed,
as we will see, unfortunately leaves out certain classes of Hamiltonians
of interest. However, we regard it as a necessary assumption which
we cannot relax in order to formulate the proof according to this
approach. Our assumptions on $H$ are thus as follows: 

~

\noindent {\bf H1} \textsl{(Kinetic terms)}{\bf:} With regards to the second partials of $H$
with respect to the momentum variables, we assume the following:

~

(a) We can make a choice of coordinates so as to diagonalise (i.e. decouple) the kinetic terms. In other words, we can choose to write $H$ in such a form that we have: 
\begin{equation}
\frac{\partial^{2}H}{\partial p_{i}\partial p_{j}}=\delta_{ij}\frac{\partial^{2}H}{\partial p_{j}^{2}}\,. \label{eq:CM_H_pipj}
\end{equation}

~

(b) Additionally, the second partials of $H$ with respect to each momentum
variable, representing the coefficients of the kinetic terms, should
be non-negative: 
\begin{equation}
\frac{\partial^{2}H}{\partial p_{j}^{2}}\geq0\,. \label{eq:CM_H_pp}
\end{equation}

~

\noindent {\bf H2} \textsl{(Mixed terms)}{\bf:} We assume that we can decouple the terms that
mix kinetic and coordinate degrees of freedom (via performing integrations
by parts, if necessary, in the action out of which the Hamiltonian
is constructed), such that $H$ can be written in a form where: 
\begin{equation}
\frac{\partial^{2}H}{\partial p_{i}\partial q_{j}}=0\,. \label{eq:CM_H_pq}
\end{equation}

~

\noindent {\bf H3} \textsl{(Potential terms)}{\bf:} We need to restrict our consideration to
Hamiltonians whose partial Hessian with respect to the coordinate
variables is positive semidefinite at the point of equilibrium (assuming
it exists), i.e. $[\partial^{2}H/\partial q_{i}\partial q_{j}]_{0}\succeq0$.
In fact we need to impose a slightly stronger (sufficient, though not strictly necessary) condition: that any of the row sums of $[\partial^{2}H/\partial q_{i}\partial q_{j}]_{0}$
are non-negative. That is to say, we assume: 
\begin{equation}
\sum_{i=1}^{N}\left(\frac{\partial^{2}H}{\partial q_{i}\partial q_{j}}\right)_{0}\geq0.\label{eq:CM_rowsum_V}
\end{equation}

We can make a few remarks about these assumptions. Firstly, H1 and H2 are manifestly satisfied for the most typically-encountered
form of the Hamiltonian in CM, 
\begin{equation}
H=\sum_{j=1}^{N}\frac{p_{j}^{2}}{2m_{j}}+V\left(q_{1},...,q_{N}\right),\label{eq:CM_typical_Hamiltonian}
\end{equation}
where $m_{j}$ are the masses associated with each degree of freedom
and $V$ is the potential (a function of only the configuration variables,
and not the momenta). Indeed, H1(a) is satisfied since we have $\partial^{2}H/\partial p_{i}\partial p_{j}=0$
unless $i=j$, regardless of $V$. For H1(b), we clearly have $\partial^{2}H/\partial p_{j}^{2}=1/m_{j}>0$
assuming masses are positive. (Theories with negative kinetic terms, i.e. ``ghosts'', are ordinarily thought of as being problematic.)
Finally, H2 holds as $\partial^{2}H/\partial p_{i}\partial q_{j}=0$ is satisfied by construction.

Secondly, For typical Hamiltonians [Eq.~\eqref{eq:CM_typical_Hamiltonian}],
H3 translates into a condition on the potential $V$, i.e. the requirement that $\sum_{i=1}^{N}(\partial^{2}V/\partial q_{i}\partial q_{j})_{0}\geq0$.
This is not necessarily satisfied in general in CM, though it is for
many systems. For example, when we have just one degree of freedom,
$N=1$, this simply means that the potential $V(q)$ is concave upward
at the point of equilibrium (thus regarded as a \emph{stable} equilibrium), i.e. $({\rm d}^{2}V(q)/{\rm d}q^{2})_{0}\geq0$, which is reasonable
to assume. As another example, for a system of harmonic oscillators
with no interactions, $V=\left(1/2\right)\sum_{j=1}^{N}m_{j}\omega_{j}^{2}q_{j}^{2}$,
we clearly have $\sum_{i=1}^{N}\partial^{2}V/\partial q_{i}\partial q_{j}=m_{j}\omega_{j}^{2}>0$
for positive masses. Indeed, even introducing interactions does not
change this so long as the couplings are mostly non-negative. (In
other words, if the negative couplings do not dominate in strength
over the positive ones.) Higher (positive) powers of the $q_{j}$
variables in $V$ are also admissible under a similar argument. However,
we can see that condition H3 [Eq.~\eqref{eq:CM_rowsum_V}] excludes
certain classes of inverse-power potentials. Most notably, it excludes
the Kepler (gravitational two-body) Hamiltonian, $H=\left(1/2m\right)(p_{1}^{2}+p_{2}^{2})-GMm/(q_{1}^{2}+q_{2}^{2})^{1/2}$,
where $q_{j}$ are the Cartesian coordinates in the orbital plane,
and $p_{j}$ the associated momenta. In this case, we have $\det([\partial^{2}H/\partial q_{i}\partial q_{j}])=-2\left(GMm\right)^{2}/(q_{1}^{2}+q_{2}^{2})^{3}<0$,
hence $[\partial^{2}H/\partial q_{i}\partial q_{j}]$ is negative
definite everywhere and therefore cannot satisfy H3 [Eq.~\eqref{eq:CM_rowsum_V}].

\subsubsection{\label{sec:3.2.4}Our proof}

\begin{figure}
\begin{centering}
\includegraphics[scale=0.35]{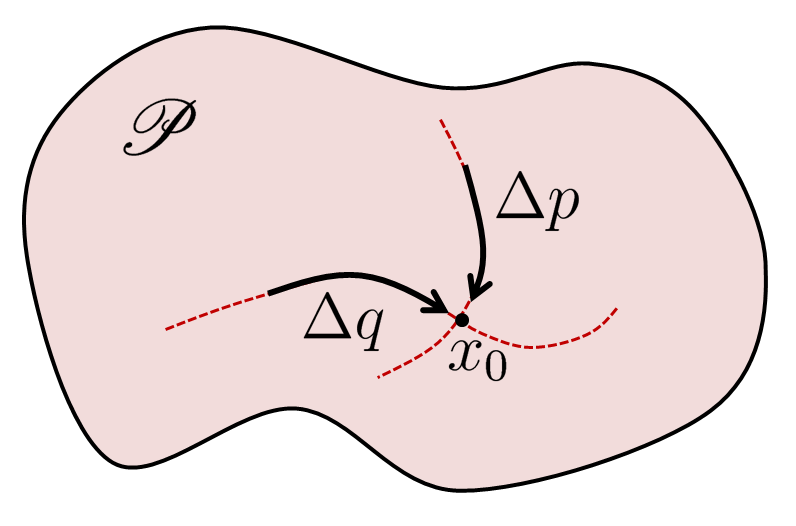}
\par\end{centering}

\protect\caption{\label{fig:perturbative}The idea of the perturbative approach is to evaluate $\dot{S}$ along different directions in phase space away from equilibrium, and arrive at a contradiction with its strict positivity.}

\end{figure}

We will now show that there cannot exist a function $S:\mathscr{P}\rightarrow\mathbb{R}$
satisfying the assumptions S1-S2 of subsubsection \ref{sec:3.2.2}
in a Hamiltonian system that obeys the assumptions H1-H3 of subsubsection
\ref{sec:3.2.3} on $H:\mathscr{P}\rightarrow\mathbb{R}$. We do this
by simply assuming that such a function exists, and we will show that
this implies a contradiction. For a pictorial representation, see Figure~\ref{fig:perturbative}.

~

\noindent$\bm{N=1}${\bf:} Let us first carry out the proof for $N=1$ degree of freedom so as to
make the argument for general $N$ easier to follow. Let $S:\mathscr{P}\rightarrow\mathbb{R}$
be any function on the configuration space $\mathscr{P}$ satisfying
assumptions S1-S2 of subsubsection \ref{sec:3.2.2}, i.e. it has an
equilibrium point and the Hessian of its time derivative is positive
definite there. We know that its time derivative at any point $x=(q,p)\in\mathscr{P}$
can be evaluated, as discussed in subsection \ref{sec:3.1}, via the
Poisson bracket: 
\begin{equation}
\dot{S}=\frac{\partial H}{\partial p}\frac{\partial S}{\partial q}-\frac{\partial H}{\partial q}\frac{\partial S}{\partial p}.\label{eq:CM_Sdot_Poisson-1}
\end{equation}
Let us now insert into this the Taylor series for each term expanded
about the equilibrium point $x_{0}=(q_{0},p_{0})$. Denoting $\Delta q=q-q_{0}$
and $\Delta p=p-p_{0}$, and using $\mathcal{O}(\Delta^{n})$ to represent $n$-th order terms in products of $\Delta q$ and $\Delta p$, we have: 
\begin{equation}
\frac{\partial H}{\partial q}=\left(\frac{\partial H}{\partial q}\right)_{0}+\left(\frac{\partial^{2}H}{\partial q^{2}}\right)_{0}\Delta q+\left(\frac{\partial^{2}H}{\partial p\partial q}\right)_{0}\Delta p+\mathcal{O}\left(\Delta^{2}\right),\label{eq:CM_dHdq-1}
\end{equation}
and similarly for the $p$ partial of $H$, while 
\begin{equation}
\frac{\partial S}{\partial q}=\left(\frac{\partial^{2}S}{\partial q^{2}}\right)_{0}\Delta q+\left(\frac{\partial^{2}S}{\partial p\partial q}\right)_{0}\Delta p+\mathcal{O}\left(\Delta^{2}\right),\label{eq:CM_dSdq-1}
\end{equation}
and similarly for the $p$ partial of $S$, where we have used the
condition S1 [Eq.~\eqref{eq:CM_Equilibrium}] which entails that
the zero-order term vanishes. Inserting all Taylor series into the
Poisson bracket [Eq.~\eqref{eq:CM_Sdot_Poisson-1}] and collecting
terms, we obtain the following result: 
\begin{equation}
\dot{S}=\!\left[\begin{array}{cc}
a & b\end{array}\right]\left[\begin{array}{c}
\Delta q\\
\Delta p
\end{array}\right]+\left[\begin{array}{cc}
\Delta q & \Delta p\end{array}\right]\left[\begin{array}{cc}
A & B\\
B & C
\end{array}\right]\left[\begin{array}{c}
\Delta q\\
\Delta p
\end{array}\right]+\mathcal{O}\left(\Delta^{3}\right)\!,\label{eq:CM_Sdot_QuadForm-1}
\end{equation}
where: 
\begin{align}
a= & \left(\frac{\partial H}{\partial p}\right)_{0}\left(\frac{\partial^{2}S}{\partial q^{2}}\right)_{0}-\left(\frac{\partial H}{\partial q}\right)_{0}\left(\frac{\partial^{2}S}{\partial q\partial p}\right)_{0},\label{eq:CM_ai-1}\\
b= & \left(\frac{\partial H}{\partial p}\right)_{0}\left(\frac{\partial^{2}S}{\partial p\partial q}\right)_{0}-\left(\frac{\partial H}{\partial q}\right)_{0}\left(\frac{\partial^{2}S}{\partial p^{2}}\right)_{0},\label{eq:CM_bi-1}
\end{align}
and: 
\begin{align}
A= & \left(\frac{\partial^{2}H}{\partial q\partial p}\right)_{0}\left(\frac{\partial^{2}S}{\partial q^{2}}\right)_{0}-\left(\frac{\partial^{2}H}{\partial q^{2}}\right)_{0}\left(\frac{\partial^{2}S}{\partial q\partial p}\right)_{0},\label{eq:CM_Aij-1}\\
B= & \frac{1}{2}\Bigg[\left(\frac{\partial^{2}H}{\partial p^{2}}\right)_{0}\left(\frac{\partial^{2}S}{\partial q^{2}}\right)_{0}-\left(\frac{\partial^{2}H}{\partial q^{2}}\right)_{0}\left(\frac{\partial^{2}S}{\partial p^{2}}\right)_{0}\Bigg],\label{eq:CM_Bij-1}\\
C= & \left(\frac{\partial^{2}H}{\partial p^{2}}\right)_{0}\left(\frac{\partial^{2}S}{\partial p\partial q}\right)_{0}-\left(\frac{\partial^{2}H}{\partial p\partial q}\right)_{0}\left(\frac{\partial^{2}S}{\partial p^{2}}\right)_{0}.\label{eq:CM_Cij-1}
\end{align}
By assumption S2, we have that $\dot{S}$ as given above [Eq.~\eqref{eq:CM_Sdot_QuadForm-1}]
is strictly positive for any $x\neq x_{0}$ in $\mathscr{P}$. In
particular, let $\delta>0$ and let us consider $\dot{S}$ [Eq.~\eqref{eq:CM_Sdot_QuadForm-1}] evaluated at the sequence of points
$\left\{ x_{n}^{\pm}\right\} _{n=1}^{\infty}$, where $x_{n}^{\pm}=(q_{0}\pm\delta/n,p_{0})$,
such that the only deviation away from equilibrium is along the direction
$\Delta q=\pm\delta/n$, with all other $\Delta p$ vanishing. Then,
for any $n$, we must have according to our expression for $\dot{S}$
[Eq.~\eqref{eq:CM_Sdot_QuadForm-1}]: 
\begin{align}
\dot{S}\left(x_{n}^{+}\right) & =a\frac{\delta}{n}+A\frac{\delta^{2}}{n^{2}}+\mathcal{O}\left(\frac{\delta^{3}}{n^{3}}\right)>0,\label{eq:CM_Sdotplus-1}\\
\dot{S}\left(x_{n}^{-}\right) & =-a\frac{\delta}{n}+A\frac{\delta^{2}}{n^{2}}+\mathcal{O}\left(\frac{\delta^{3}}{n^{3}}\right)>0.\label{eq:CM_Sdotminus-1}
\end{align}
Taking the $n\rightarrow\infty$ limit of the first inequality implies
$a\geq0$, while doing the same for the second inequality implies
$a\leq0$. Hence $a=0$. A similar argument (using $\Delta p=\pm\delta/n$)
implies $b=0$. Thus, $S$ needs to satisfy the constraints 
\begin{align}
0= & \left(\frac{\partial H}{\partial p}\right)_{0}\left(\frac{\partial^{2}S}{\partial q^{2}}\right)_{0}-\left(\frac{\partial H}{\partial q}\right)_{0}\left(\frac{\partial^{2}S}{\partial q\partial p}\right)_{0},\label{eq:CM_ai0-1}\\
0= & \left(\frac{\partial H}{\partial p}\right)_{0}\left(\frac{\partial^{2}S}{\partial p\partial q}\right)_{0}-\left(\frac{\partial H}{\partial q}\right)_{0}\left(\frac{\partial^{2}S}{\partial p^{2}}\right)_{0},\label{eq:CM_bi0-1}
\end{align}
and this leaves us with 
\begin{equation}
\dot{S}=\left[\begin{array}{cc}
\Delta q & \Delta p\end{array}\right]\left[\begin{array}{cc}
A & B\\
B & C
\end{array}\right]\left[\begin{array}{c}
\Delta q\\
\Delta p
\end{array}\right]+\mathcal{O}\left(\Delta^{3}\right).\label{eq:CM_Sdot_QuadForm2-1}
\end{equation}

Now, imposing the Hamiltonian assumptions H1(a) and H2 {[}eqs. (\ref{eq:CM_H_pipj})
and (\ref{eq:CM_H_pq}) respectively{]} simplifies $A$ and $C$,
from the above {[}eqs. (\ref{eq:CM_Aij-1}) and (\ref{eq:CM_Bij-1})
respectively{]} to: 
\begin{align}
A=\, & -\left(\frac{\partial^{2}H}{\partial q^{2}}\right)_{0}\left(\frac{\partial^{2}S}{\partial q\partial p}\right)_{0},\label{eq:CM_Aij_simplified-1}\\
C=\, & \left(\frac{\partial^{2}H}{\partial p^{2}}\right)_{0}\left(\frac{\partial^{2}S}{\partial p\partial q}\right)_{0}.\label{eq:CM_Cij_simplified-1}
\end{align}
Positive-definiteness of $(\mathbf{Hess}(\dot{S}))_{0}$ (assumption
S2) implies that the quadratic form above [Eq.~\eqref{eq:CM_Sdot_QuadForm2-1}]
should be positive definite. This means that we cannot have $(\partial^{2}H/\partial p^{2})_{0}=0$,
since then $C$ would not be strictly positive and we would get a
contradiction. This, combined with assumption H1(b) [Eq.~\eqref{eq:CM_H_pp}],
implies that $(\partial^{2}H/\partial p^{2})_{0}>0$. This in combination
with $C>0$ means that $(\partial^{2}S/\partial p\partial q)_{0}>0$.
But $A>0$ also, in order to have positive-definiteness of the quadratic
form [Eq.~\eqref{eq:CM_Sdot_QuadForm2-1}], and this combined
with assumption H3 [Eq.~\eqref{eq:CM_rowsum_V}], i.e. $(\partial^{2}H/\partial q^{2})_{0}\geq0$,
implies $(\partial^{2}S/\partial p\partial q)_{0}<0$. Thus we get
a contradiction, and so no such function $S$ exists.

~

\noindent{\bf General} $\bm{N}${\bf:} The extension of the proof to general $N$ follows similar lines,
though with a few added subtleties. Let us now proceed with it. As
before, suppose $S:\mathscr{P}\rightarrow\mathbb{R}$ is any function
on $\mathscr{P}$ satisfying S1-S2. Its time derivative at any point
$x=(q_{1},...,q_{N},p_{1},...,p_{N})\in\mathscr{P}$ can be evaluated
via the Poisson bracket: 
\begin{equation}
\dot{S}=\sum_{k=1}^{N}\left(\frac{\partial H}{\partial p_{k}}\frac{\partial S}{\partial q_{k}}-\frac{\partial H}{\partial q_{k}}\frac{\partial S}{\partial p_{k}}\right).\label{eq:CM_Sdot_Poisson}
\end{equation}
Let us now insert into this the Taylor series for each term expanded
about the equilibrium point $x_{0}=((q_{0})_{1},...,(q_{0})_{N},(p_{0})_{1},...,(p_{0})_{1})$.
Denoting $\Delta q_{i}=q_{i}-(q_{0})_{i}$ and $\Delta p_{i}=p_{i}-(p_{0})_{i}$, and using $\mathcal{O}(\Delta^{n})$ to represent $n$-th order terms in products of $\Delta q_{i}$ and $\Delta p_{i}$,
we have: 
\begin{multline}
\frac{\partial H}{\partial q_{k}}=\left(\frac{\partial H}{\partial q_{k}}\right)_{0}\\
+\sum_{i=1}^{N}\Bigg[\left(\frac{\partial^{2}H}{\partial q_{i}\partial q_{k}}\right)_{0}\Delta q_{i}+\left(\frac{\partial^{2}H}{\partial p_{i}\partial q_{k}}\right)_{0}\Delta p_{i}\Bigg]+\mathcal{O}\left(\Delta^{2}\right),\label{eq:CM_dHdq}
\end{multline}
and similarly for the $p_{k}$ partial of $H$, while 
\begin{equation}
\frac{\partial S}{\partial q_{k}}\!=\sum_{i=1}^{N}\left[\left(\frac{\partial^{2}S}{\partial q_{i}\partial q_{k}}\right)_{0}\Delta q_{i}+\left(\frac{\partial^{2}S}{\partial p_{i}\partial q_{k}}\right)_{0}\Delta p_{i}\right]+\mathcal{O}\left(\Delta^{2}\right)\!,\label{eq:CM_dSdq}
\end{equation}
and similarly for the $p_{k}$ partial of $S$, where we have used
the condition S1 [Eq.~\eqref{eq:CM_Equilibrium}] which entails
that the zero-order term vanishes. Inserting all Taylor series into
the Poisson bracket [Eq.~\eqref{eq:CM_Sdot_Poisson}] and collecting
terms, we obtain the following result: 
\begin{multline}
\dot{S}=\left[\begin{array}{cc}
\mathbf{a}^{{\rm T}} & \mathbf{b}^{{\rm T}}\end{array}\right]\left[\begin{array}{c}
\Delta q_{1}\\
\vdots\\
\Delta p_{N}
\end{array}\right]\\
+\left[\begin{array}{ccc}
\Delta q_{1} & \cdots & \Delta p_{N}\end{array}\right]\left[\begin{array}{cc}
\mathbf{A} & \mathbf{B}\\
\mathbf{B}^{{\rm T}} & \mathbf{C}
\end{array}\right]\left[\begin{array}{c}
\Delta q_{1}\\
\vdots\\
\Delta p_{N}
\end{array}\right]+\mathcal{O}\left(\Delta^{3}\right),\label{eq:CM_Sdot_QuadForm}
\end{multline}
where we have the following components for the $N$-dimensional vectors:
\begin{equation}
a_{i}=\sum_{k=1}^{N}\Bigg[\left(\frac{\partial H}{\partial p_{k}}\right)_{0}\!\!\left(\frac{\partial^{2}S}{\partial q_{i}\partial q_{k}}\right)_{0}\!\!-\left(\frac{\partial H}{\partial q_{k}}\right)_{0}\!\!\left(\frac{\partial^{2}S}{\partial q_{i}\partial p_{k}}\right)_{0}\!\!\Bigg],\label{eq:CM_ai}
\end{equation}
and
\begin{equation}
b_{i}=\sum_{k=1}^{N}\Bigg[\left(\frac{\partial H}{\partial p_{k}}\right)_{0}\!\!\left(\frac{\partial^{2}S}{\partial p_{i}\partial q_{k}}\right)_{0}\!\!-\left(\frac{\partial H}{\partial q_{k}}\right)_{0}\!\!\left(\frac{\partial^{2}S}{\partial p_{i}\partial p_{k}}\right)_{0}\!\!\Bigg],\label{eq:CM_bi}
\end{equation}
and for the $N\times N$ matrices:
\begin{widetext}
\begin{align}
A_{ij}=&\frac{1}{2}\sum_{k=1}^{N}\Bigg[\left(\frac{\partial^{2}H}{\partial q_{i}\partial p_{k}}\right)_{0}\!\!\left(\frac{\partial^{2}S}{\partial q_{j}\partial q_{k}}\right)_{0}\!\!+\left(\frac{\partial^{2}H}{\partial q_{j}\partial p_{k}}\right)_{0}\!\!\left(\frac{\partial^{2}S}{\partial q_{i}\partial q_{k}}\right)_{0}\!\!-\left(\frac{\partial^{2}H}{\partial q_{i}\partial q_{k}}\right)_{0}\!\!\left(\frac{\partial^{2}S}{\partial q_{j}\partial p_{k}}\right)_{0}\!\!-\left(\frac{\partial^{2}H}{\partial q_{j}\partial q_{k}}\right)_{0}\!\!\left(\frac{\partial^{2}S}{\partial q_{i}\partial p_{k}}\right)_{0}\!\!\Bigg],\label{eq:CM_Aij}\\
B_{ij}=&\frac{1}{2}\sum_{k=1}^{N}\Bigg[ \left(\frac{\partial^{2}H}{\partial q_{i}\partial p_{k}}\right)_{0}\!\!\left(\frac{\partial^{2}S}{\partial p_{j}\partial q_{k}}\right)_{0}\!\!+\left(\frac{\partial^{2}H}{\partial p_{j}\partial p_{k}}\right)_{0}\!\!\left(\frac{\partial^{2}S}{\partial q_{i}\partial q_{k}}\right)_{0}\!\! -\left(\frac{\partial^{2}H}{\partial q_{i}\partial q_{k}}\right)_{0}\!\!\left(\frac{\partial^{2}S}{\partial p_{j}\partial p_{k}}\right)_{0}\!\!-\left(\frac{\partial^{2}H}{\partial p_{j}\partial q_{k}}\right)_{0}\!\!\left(\frac{\partial^{2}S}{\partial q_{i}\partial p_{k}}\right)_{0}\!\!\Bigg],\label{eq:CM_Bij}\\
C_{ij}=&\frac{1}{2}\sum_{k=1}^{N}\Bigg[\left(\frac{\partial^{2}H}{\partial p_{i}\partial p_{k}}\right)_{0}\!\!\left(\frac{\partial^{2}S}{\partial p_{j}\partial q_{k}}\right)_{0}\!\!+\left(\frac{\partial^{2}H}{\partial p_{j}\partial p_{k}}\right)_{0}\!\!\left(\frac{\partial^{2}S}{\partial p_{i}\partial q_{k}}\right)_{0}\!\! -\left(\frac{\partial^{2}H}{\partial p_{i}\partial q_{k}}\right)_{0}\!\!\left(\frac{\partial^{2}S}{\partial p_{j}\partial p_{k}}\right)_{0}\!\!-\left(\frac{\partial^{2}H}{\partial p_{j}\partial q_{k}}\right)_{0}\!\!\left(\frac{\partial^{2}S}{\partial p_{i}\partial p_{k}}\right)_{0}\!\!\Bigg].\label{eq:CM_Cij}
\end{align}
\end{widetext}
By assumption S2, we have that $\dot{S}$ as given above [Eq.~\eqref{eq:CM_Sdot_QuadForm}]
is strictly positive for any $x\neq x_{0}$ in $\mathscr{P}$. In
particular, let $\delta>0$ and let us consider $\dot{S}$ [Eq.~\eqref{eq:CM_Sdot_QuadForm}] evaluated at the sequence of points
$\left\{ x_{n}^{\pm}\right\} _{n=1}^{\infty}$, where $x_{n}^{\pm}=((q_{0})_{1},...,(q_{0})_{l}\pm\delta/n,...,(q_{0})_{N},(p_{0})_{1},...,(p_{0})_{1})$,
for any $l$, such that the only deviation away from equilibrium is
along the direction $\Delta q_{l}=\pm\delta/n$, with all other $\Delta q_{i}$
and $\Delta p_{i}$ vanishing. Then, for any $n$, we must have according
to the above expression for $\dot{S}$ [Eq.~\eqref{eq:CM_Sdot_QuadForm}]:
\begin{align}
\dot{S}\left(x_{n}^{+}\right) & =a_{l}\frac{\delta}{n}+A_{ll}\frac{\delta^{2}}{n^{2}}+\mathcal{O}\left(\frac{\delta^{3}}{n^{3}}\right)>0,\label{eq:CM_Sdotplus}\\
\dot{S}\left(x_{n}^{-}\right) & =-a_{l}\frac{\delta}{n}+A_{ll}\frac{\delta^{2}}{n^{2}}+\mathcal{O}\left(\frac{\delta^{3}}{n^{3}}\right)>0.\label{eq:CM_Sdotminus}
\end{align}
Taking the $n\rightarrow\infty$ limit of the first inequality implies
$a_{l}\geq0$, while doing the same for the second inequality implies
$a_{l}\leq0$. Hence $a_{l}=0$. Since $l$ is arbitrary, this means
that $a_{i}=0$, $\forall i$. A similar argument (using $\Delta p_{l}=\pm\delta/n$)
implies $b_{i}=0$, $\forall i$. Thus, $S$ needs to satisfy the
constraints 
\begin{align}
0= & \sum_{k=1}^{N}\Bigg[\left(\frac{\partial H}{\partial p_{k}}\right)_{0}\left(\frac{\partial^{2}S}{\partial q_{i}\partial q_{k}}\right)_{0}-\left(\frac{\partial H}{\partial q_{k}}\right)_{0}\left(\frac{\partial^{2}S}{\partial q_{i}\partial p_{k}}\right)_{0}\Bigg],\label{eq:CM_ai0}\\
0= & \sum_{k=1}^{N}\Bigg[\left(\frac{\partial H}{\partial p_{k}}\right)_{0}\left(\frac{\partial^{2}S}{\partial p_{i}\partial q_{k}}\right)_{0}-\left(\frac{\partial H}{\partial q_{k}}\right)_{0}\left(\frac{\partial^{2}S}{\partial p_{i}\partial p_{k}}\right)_{0}\Bigg],\label{eq:CM_bi0}
\end{align}
and this leaves us with 
\begin{equation}
\dot{S}=\left[\begin{array}{ccc}
\Delta q_{1} & \cdots & \Delta p_{N}\end{array}\right]\left[\begin{array}{cc}
\mathbf{A} & \mathbf{B}\\
\mathbf{B}^{{\rm T}} & \mathbf{C}
\end{array}\right]\left[\begin{array}{c}
\Delta q_{1}\\
\vdots\\
\Delta p_{N}
\end{array}\right]+\mathcal{O}\left(\Delta^{3}\right).\label{eq:CM_Sdot_QuadForm2}
\end{equation}

Now, imposing the Hamiltonian assumptions H1(a) and H2 {[}eqs. (\ref{eq:CM_H_pipj})
and (\ref{eq:CM_H_pq}) respectively{]} simplifies $\mathbf{A}$ and
$\mathbf{C}$, from the above {[}eqs. (\ref{eq:CM_Aij}) and (\ref{eq:CM_Cij})
respectively{]} to: 
\begin{align}
A_{ij}=\, & -\frac{1}{2}\sum_{k=1}^{N}\Bigg[\left(\frac{\partial^{2}H}{\partial q_{i}\partial q_{k}}\right)_{0}\left(\frac{\partial^{2}S}{\partial q_{j}\partial p_{k}}\right)_{0}\\
 & \quad\quad\quad\quad +\left(\frac{\partial^{2}H}{\partial q_{j}\partial q_{k}}\right)_{0}\left(\frac{\partial^{2}S}{\partial q_{i}\partial p_{k}}\right)_{0}\Bigg],\label{eq:CM_Aij_simplified}\\
C_{ij}=\, & \frac{1}{2}\Bigg[\left(\frac{\partial^{2}H}{\partial p_{i}^{2}}\right)_{0}\!\!\left(\frac{\partial^{2}S}{\partial p_{j}\partial q_{i}}\right)_{0}\!\!+\left(\frac{\partial^{2}H}{\partial p_{j}^{2}}\right)_{0}\!\!\left(\frac{\partial^{2}S}{\partial p_{i}\partial q_{j}}\right)_{0}\!\!\Bigg].\label{eq:CM_Cij_simplified}
\end{align}
Positive-definiteness of $(\mathbf{Hess}(\dot{S}))_{0}$ implies that
the quadratic form above [Eq.~\eqref{eq:CM_Sdot_QuadForm2}] should
be positive definite. This means that we cannot have $(\partial^{2}H/\partial p_{j}^{2})_{0}=0$,
$\forall j$, since then $\mathbf{C}$ would not be positive definite
and we would get a contradiction. This, combined with assumption H1(b)
[Eq.~\eqref{eq:CM_H_pp}], implies that $(\partial^{2}H/\partial p_{j}^{2})_{0}>0$,
$\forall j$. Moreover, we also have: 
\begin{equation}
\sum_{i,j=1}^{N}C_{ij}=\sum_{i,j=1}^{N}\left(\frac{\partial^{2}H}{\partial p_{j}^{2}}\right)_{0}\left(\frac{\partial^{2}S}{\partial p_{i}\partial q_{j}}\right)_{0}>0.\label{eq:CM_Cij_posdef}
\end{equation}
The reason for this is easily seen by noting that positive-definiteness of $\mathbf{C}$, by definition, means that its product with any nonzero
vector and its transpose should be positive, i.e. $\mathbf{z}^{{\rm T}}\mathbf{C}\mathbf{z}>0$
for any nonzero vector $\mathbf{z}$; in particular, $\mathbf{z}=(1,1,...,1)^{{\rm T}}$
achieves the above inequality [Eq.~\eqref{eq:CM_Cij_posdef}].
But then, let us consider $\sum_{i,j=1}^{N}A_{ij}$. Positive-definiteness
of $(\mathbf{Hess}(\dot{S}))_{0}$ (i.e. of the quadratic form [Eq.~\eqref{eq:CM_Sdot_QuadForm}]) implies, just as in the case of $\mathbf{C}$,
that $\sum_{i,j=1}^{N}A_{ij}>0$, or 
\begin{equation}
\sum_{i,j=1}^{N}(-A_{ij})<0\label{CM_Aij_ineq}.
\end{equation}
At the same time, we have: 
\begin{equation}
\sum_{i,j=1}^{N}\left(-A_{ij}\right)=\sum_{i,j,k=1}^{N}\left(\frac{\partial^{2}H}{\partial q_{i}\partial q_{k}}\right)_{0}\left(\frac{\partial^{2}S}{\partial q_{j}\partial p_{k}}\right)_{0}.\label{CM_Aij_eq}
\end{equation}
Taking the minimum over the $k$ index in the term with the $H$ partials,
\begin{equation}
\sum_{i,j=1}^{N}\left(-A_{ij}\right)\geq\sum_{i,j,k=1}^{N}\left[\min_{1\leq l\leq N}\left(\frac{\partial^{2}H}{\partial q_{i}\partial q_{l}}\right)_{0}\right]\left(\frac{\partial^{2}S}{\partial q_{j}\partial p_{k}}\right)_{0},\label{eq:CM_-Aij_1}
\end{equation}
This means that the sums can be separated, and after relabelling,
the above [Eq.~\eqref{eq:CM_-Aij_1}] becomes: 
\begin{equation}
\sum_{i,j=1}^{N}\left(-A_{ij}\right)\geq\left[\min_{1\leq l\leq N}\sum_{k=1}^{N}\left(\frac{\partial^{2}H}{\partial q_{k}\partial q_{l}}\right)_{0}\right]\sum_{i,j=1}^{N}\left(\frac{\partial^{2}S}{\partial p_{i}\partial q_{j}}\right)_{0}.\label{eq:CM_-Aij_2}
\end{equation}
Now, insert the identity $1=(\partial^{2}H/\partial p_{j}^{2})_{0}/(\partial^{2}H/\partial p_{j}^{2})_{0}$
into the $i,j$ sum, and maximise over the denominator to get: 
\begin{widetext}
\begin{align}
\sum_{i,j=1}^{N}\left(-A_{ij}\right)\geq & \left[\min_{1\leq l\leq N}\sum_{k=1}^{N}\left(\frac{\partial^{2}H}{\partial q_{k}\partial q_{l}}\right)_{0}\right]\sum_{i,j=1}^{N}\frac{\left(\partial^{2}H/\partial p_{j}^{2}\right)_{0}}{\left(\partial^{2}H/\partial p_{j}^{2}\right)_{0}}\left(\frac{\partial^{2}S}{\partial p_{i}\partial q_{j}}\right)_{0}\label{eq:CM_-Aij_3}\\
\geq & \left[\min_{1\leq l\leq N}\sum_{k=1}^{N}\left(\frac{\partial^{2}H}{\partial q_{k}\partial q_{l}}\right)_{0}\right]\sum_{i,j=1}^{N}\left[\max_{1\leq m\leq N}\left(\frac{\partial^{2}H}{\partial p_{m}^{2}}\right)_{0}\right]^{-1}\left(\frac{\partial^{2}H}{\partial p_{j}^{2}}\right)_{0}\left(\frac{\partial^{2}S}{\partial p_{i}\partial q_{j}}\right)_{0}\label{eq:CM_-Aij_4}\\
= & \left\{ \left[\min_{1\leq l\leq N}\sum_{k=1}^{N}\left(\frac{\partial^{2}H}{\partial q_{k}\partial q_{l}}\right)_{0}\right]\left[\max_{1\leq m\leq N}\left(\frac{\partial^{2}H}{\partial p_{m}^{2}}\right)_{0}\right]^{-1}\right\} \sum_{i,j=1}^{N}C_{ij}\label{eq:CM_-Aij_5}\\
\geq & \,0,\label{eq:CM_-Aij_6}
\end{align}
\end{widetext}
since the term in curly brackets is non-negative (because of assumption
H3 on the Hamiltonian), and we had earlier $\sum_{i,j=1}^{N}C_{ij}>0$.
But we also had $\sum_{i,j=1}^{N}(-A_{ij})<0$. Hence we get a contradiction.
Therefore, no such function $S$ exists. This concludes our proof.

\subsection{\label{sec:3.3}Topological approach}

We now turn to the topological approach to answering Problem I in CM.
First we review
the basic ideas of Olsen's line of argumentation \cite{olsen_classical_1993}, then we discuss their connections with the periodicity of phase space orbits.

\subsubsection{\label{sec:3.3.1}Review of Olsen's proof}

\begin{figure}
\begin{centering}
\includegraphics[scale=0.35]{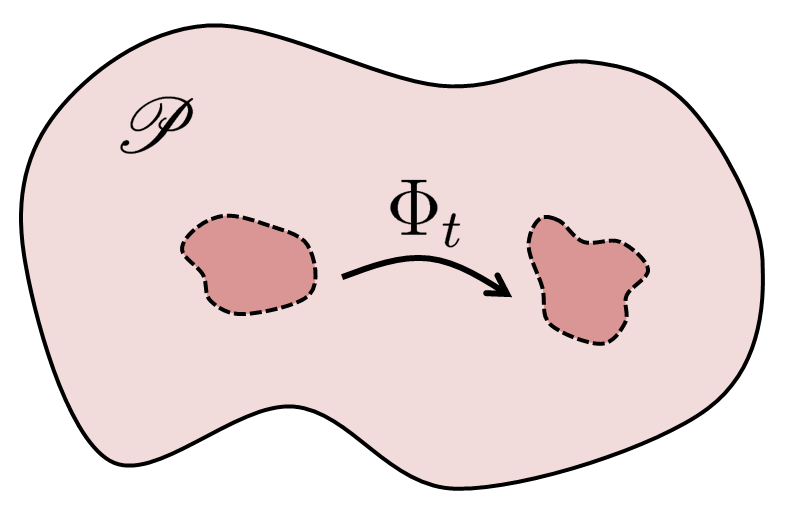}
\par\end{centering}

\protect\caption{\label{fig:topological}The topological approach relies on phase space compactness and Liouville's theorem, i.e. the fact that the Hamiltonian flow is volume-preserving.}

\end{figure}

The assumptions made on $S:\mathscr{P}\rightarrow\mathbb{R}$ are
in this case not as strict as in the perturbative approach. See Figure \ref{fig:topological} for a pictorial representation.

In effect, we simply
need to assume that $S$ is nondecreasing along trajectories, which
are confined to an invariant closed space $\mathscr{P}$. Under these conditions, Olsen furnishes two proofs \cite{olsen_classical_1993} for why $S$ is necessarily a constant. In the first one, the
essential idea is that the volume integral of $S$ in $\mathscr{P}$ can be written after a change of variables
as 
\begin{equation}
\int_{\mathscr{P}}\bm{\Omega}\, S=\int_{\mathscr{P}}\bm{\Omega}\, \left(S\circ\Phi_{t}\right),\label{eq:CM_Olsen_pf}
\end{equation}
owing to the fact that $\mathscr{P}$ is left invariant by the Hamiltonian
flow $\Phi_{t}$ generated by the Hamiltonian vector field $\bm{X}_{H}$,
and that $\mathcal{L}_{\bm{X}_{H}}\bm{\Omega}=0$. Because the above
expression [Eq.~\eqref{eq:CM_Olsen_pf}] is time-independent,
$S$ must be time-independent, hence constant along trajectories. The second proof (based on the same assumptions) is rather more technical, but relies also basically on topological ideas; in fact, it is more related to the Poincar\'{e} recurrence property \cite{luis_barreira_poincare_2006}.

We can make a few remarks. Firstly, there is in this case no requirement on the specific form of the Hamiltonian
function $H:\mathscr{P\rightarrow\mathbb{R}}$. In fact, $H$ can
even contain explicit dependence on time and the proof still holds.

Secondly, the essential ingredient here is the compactness of the phase space
$\mathscr{P}$. Indeed, even in Poincar\'{e}'s original recurrence theorem \cite{hp1890am} (as per our discussion in appendix \ref{sec:A}), the only necessary assumptions were also phase space compactness and invariance along with Liouville's theorem. 

\subsubsection{\label{sec:3.3.2}Periodicity in phase space}

Even more can be said about the connection between phase space compactness and the recurrence of orbits than the Poincar\'{e} recurrence theorem. There are recent theorems in symplectic geometry which show that exact periodicity of orbits can exist in compact phase spaces.

For example, let us assume the Hamiltonian is of typical form [Eq.~\eqref{eq:CM_typical_Hamiltonian}]. Then, there is a theorem \cite{hofer_symplectic_2011} which
states that for a compact configuration space $\mathscr{Q}$, we have
periodic solutions of $\bm{X}_{H}$. In fact, it was even shown \cite{suhr_linking_2016} that
we have periodic solutions provided certain conditions on the potential
$V$ are satisfied and $\mathscr{Q}$ just needs to have bounded geometry
(i.e. to be geodesically complete and to have the scalar curvature
and derivative thereof bounded).

Thus, under the assumption of compactness or any other condition which
entails closed orbits, we cannot have a function which behaves like
entropy in this sense for a very simple reason. Assume $S:\mathscr{P}\rightarrow\mathbb{R}$
is nondecreasing along trajectories and let us consider an orbit $\gamma:\mathbb{R}\rightarrow\mathscr{P}$
in phase space (satisfying ${\rm d}\gamma\left(t\right)/{\rm d}t=\bm{X}_{H}(\gamma\left(t\right))$)
which is closed. This means that for any $x\in\mathscr{P}$ on the
orbit, there exist $t_{0},T\in\mathbb{R}$ such that $x=\gamma(t_{0})=\gamma(t_{0}+T)$.
Hence, we have $S(x)=S(\gamma(t_{0}))=S(\gamma(t_{0}+T))=S(x)$, so $S$ is
constant along the orbit and therefore cannot behave like entropy.

\section{\label{sec:GR}Entropy theorems in general relativity}

We now turn to addressing the question of why these theorems do not carry over from CM to GR.

\subsection{\label{sec:4.1}Setup}

Let $\mathscr{M}$ be a 4-manifold with a Lorentzian metric $\bm{g}$
of signature $\left(-,+,+,+\right)$ and having a metric-compatible
derivative operator $\nabla$. We choose a coordinate system where
$t$ denotes the time coordinate, and $\bm{t}$ is the time flow vector
field on $\mathscr{M}$. We label $\Sigma$ the $t={\rm const}.$
spacelike 3-surfaces foliating $\mathscr{M}$, such that $\mathscr{M}=\mathbb{R}\times\Sigma$.
On $\Sigma$, we have an induced 3-dimensional Riemannian metric $\bm{h}$,
with a metric-compatible derivative operator $D$. Its conjugate momentum
is $\bm{\pi}=\sqrt{h}\left(\bm{K}-K\bm{h}\right),$ where $h={\rm det}\left(\bm{h}\right)$,
$\bm{K}$ is the extrinsic curvature of $\Sigma$, and $K={\rm tr}\left(\bm{K}\right)$.
Moreover, let $N$ and $\bm{N}$, respectively, be the lapse function
and shift vector of $\bm{g}$.

Consider any field theory on $\mathscr{M}$, in this case described
by an infinite-dimensional phase space $\mathscr{P}$. Let us write,
in general, any point in phase space as $\left(\varphi,\pi\right)\in\mathscr{P}$,
where $\pi=\left\{ \pi_{A}\left(x\right)\right\} $ is the set of
momenta canonically conjugate to the fields $\varphi=\left\{ \varphi_{A}\left(x\right)\right\} $,
and $A$ is a general (possibly multi-) index for the fields. For
any functional $F:\mathscr{P}\rightarrow\mathbb{R}$, we can compute
its time derivative $\dot{F}=\mathcal{L}_{\bm{t}}F=\left\{F,H\right\}$ via a suitably-defined
Poisson bracket with the Hamiltonian $H:\mathscr{P}\rightarrow\mathbb{R}$,
which takes an analogous form as that in CM but expressed in terms
of functional derivatives \cite{bojowald_canonical_2011}: 
\begin{equation}
\dot{F}=\int_{\Sigma}{\rm d}^{3}x\sum_{A}\left(\frac{\delta H}{\delta\pi_{A}\left(x\right)}\frac{\delta F}{\delta\varphi_{A}\left(x\right)}-\frac{\delta H}{\delta\varphi_{A}\left(x\right)}\frac{\delta F}{\delta\pi_{A}\left(x\right)}\right).\label{eq:GR_Poisson_bracket_general}
\end{equation}
In particular, we will have canonical equations of motion $\dot{\varphi}_{A}=\{\varphi_{A},H\}$
and $\dot{\pi}_{A}=\{\pi_{A},H\}$.

For defining and working with differential forms on $\mathscr{P}$,
we also require a notion of an exterior derivative. Because we are
dealing with field theories, we must use the functional exterior derivative
$\delta$ (see \cite{crnkovic_symplectic_1987,crnkovic_covariant_1989} for more details): for example, $\delta\varphi_{A}\left(x\right)$ is a one-form
on $\mathscr{P}$, and so for any functional (zero-form) $\Gamma\left[\varphi_{A}\left(x\right)\right]$
on $\mathscr{P}$, for example, we have that the action of $\delta$
is given by $\delta\Gamma\left[\varphi_{A}\left(x\right)\right]=\int_{\Sigma}{\rm d}^{3}x\left(\delta\Gamma/\delta\varphi_{A}\left(x\right)\right)\delta\varphi_{A}\left(x\right)$.
This can then be extended to also define $p$-forms.

\subsection{\label{sec:4.2}Perturbative approach}

We wish to investigate under what conditions the CM no-entropy proof
of subsection \ref{sec:3.2} transfers over to field theories in curved
spacetime. To this effect, we consider the equivalent setup: broadly
speaking, we ask whether there exists a phase space functional $S:\mathscr{P}\rightarrow\mathbb{R}$
which is increasing in time everywhere except at an ``equilibrium''
configuration. In particular, we use the following two entropy conditions
in analogy with those of subsubsection \ref{sec:3.2.3} in CM: 

~

\noindent{\bf S1} \textsl{(Existence of equilibrium)}{\bf:} We assume there exists a point
$x_{0}=\left(\mathring{\varphi},\mathring{\pi}\right)\in\mathscr{P}$,
where $S$ is stationary, and (to simplify the analysis) $H$ is stationary
as well: 
\begin{equation}
\frac{\delta S\left[\mathring{\varphi},\mathring{\pi}\right]}{\delta\mathring{\varphi}_{A}\left(x\right)}=\frac{\delta S\left[\mathring{\varphi},\mathring{\pi}\right]}{\delta\mathring{\pi}_{A}\left(x\right)}=0=\frac{\delta H\left[\mathring{\varphi},\mathring{\pi}\right]}{\delta\mathring{\varphi}_{A}\left(x\right)}=\frac{\delta H\left[\mathring{\varphi},\mathring{\pi}\right]}{\delta\mathring{\pi}_{A}\left(x\right)}.\label{eq:GR_stationarity_assumption_general}
\end{equation}
This implies $\dot{S}[\mathring{\varphi},\mathring{\pi}]=0=\dot{H}[\mathring{\varphi},\mathring{\pi}]$. 

~

\noindent{\bf S2} \textsl{(Second law of thermodynamics)}{\bf:} We assume that the Hessian
of $\dot{S}$ is positive definite at equilibrium, i.e. $\mathbf{Hess}(\dot{S}[\mathring{\varphi},\mathring{\pi}])\succ0$.
This is a sufficient condition to ensure that $\dot{S}>0$ in $\mathscr{P}\backslash x_{0}$,
and $\dot{S}=0$ at $x_{0}$.

~

We then follow the same procedure as in subsubsection \ref{sec:3.2.4}:
we insert into the Poisson bracket 
\begin{equation}
\dot{S}\!=\!\int_{\Sigma}\!{\rm d}^{3}x\sum_{A}\left(\frac{\delta H\left[\varphi,\pi\right]}{\delta\pi_{A}\left(x\right)}\frac{\delta S\left[\varphi,\pi\right]}{\delta\varphi_{A}\left(x\right)}-\frac{\delta H\left[\varphi,\pi\right]}{\delta\varphi_{A}\left(x\right)}\frac{\delta S\left[\varphi,\pi\right]}{\delta\pi_{A}\left(x\right)}\right)\label{eq:GR_poisson_bracket_S_general}
\end{equation}
the functional Taylor series \cite{dreizler_density_2011} for each term about $\left(\mathring{\varphi},\mathring{\pi}\right)$,
denoting $\Delta\varphi_{A}\left(x\right)=\varphi_{A}\left(x\right)-\mathring{\varphi}\left(x\right)$ and
$\Delta\pi_{A}\left(x\right)=\pi_{A}\left(x\right)-\mathring{\pi}\left(x\right)$:
\begin{align}
\frac{\delta H\left[\varphi,\pi\right]}{\delta\pi_{A}\left(x\right)}= & \frac{\delta H\left[\mathring{\varphi},\mathring{\pi}\right]}{\delta\mathring{\pi}_{A}\left(x\right)}\nonumber\\
 & +\int_{\Sigma}{\rm d}^{3}y\sum_{B}\Bigg\{ \frac{\delta^{2}H\left[\mathring{\varphi},\mathring{\pi}\right]}{\delta\mathring{\varphi}_{B}\left(y\right)\delta\mathring{\pi}_{A}\left(x\right)}\Delta\mathring{\varphi}_{B}\left(y\right)\nonumber\\
 & +\frac{\delta^{2}H\left[\mathring{\varphi},\mathring{\pi}\right]}{\delta\mathring{\pi}_{B}\left(y\right)\delta\mathring{\pi}_{A}\left(x\right)}\Delta\mathring{\pi}_{B}\left(y\right)\Bigg\} +\mathcal{O}\left(\Delta^{2}\right),\label{eq:GR_taylor_series_general}
\end{align}
and similarly for the other terms. Then we apply S1 in this case [Eq.~\eqref{eq:GR_stationarity_assumption_general}], which makes all
zero-order terms vanish. Finally, the Poisson bracket in this case
[Eq.~\eqref{eq:GR_poisson_bracket_S_general}] becomes: 
\begin{widetext}
\begin{align}
\dot{S}=\!\!\int_{\Sigma}\!\!{\rm d}^{3}x\int_{\Sigma}\!\!{\rm d}^{3}y\int_{\Sigma}\!\!{\rm d}^{3}z\!\!\!\sum_{A,B,C}\!\!\Bigg\{ & \Bigg[\frac{\delta^{2}H\left[\mathring{\varphi},\mathring{\pi}\right]}{\delta\mathring{\varphi}_{B}\left(y\right)\delta\mathring{\pi}_{A}\left(x\right)}\frac{\delta^{2}S\left[\mathring{\varphi},\mathring{\pi}\right]}{\delta\mathring{\varphi}_{C}\left(z\right)\delta\mathring{\varphi}_{A}\left(x\right)}-\frac{\delta^{2}H\left[\mathring{\varphi},\mathring{\pi}\right]}{\delta\mathring{\varphi}_{B}\left(y\right)\delta\mathring{\varphi}_{A}\left(x\right)}\frac{\delta^{2}S\left[\mathring{\varphi},\mathring{\pi}\right]}{\delta\mathring{\varphi}_{C}\left(z\right)\delta\mathring{\pi}_{A}\left(x\right)}\Bigg]\Delta\varphi_{B}\left(y\right)\Delta\varphi_{C}\left(z\right)\nonumber \\
+ & \Bigg[\frac{\delta^{2}H\left[\mathring{\varphi},\mathring{\pi}\right]}{\delta\mathring{\varphi}_{B}\left(y\right)\delta\mathring{\pi}_{A}\left(x\right)}\frac{\delta^{2}S\left[\mathring{\varphi},\mathring{\pi}\right]}{\delta\mathring{\pi}_{C}\left(z\right)\delta\mathring{\varphi}_{A}\left(x\right)}+\frac{\delta^{2}H\left[\mathring{\varphi},\mathring{\pi}\right]}{\delta\mathring{\pi}_{C}\left(z\right)\delta\mathring{\pi}_{A}\left(x\right)}\frac{\delta^{2}S\left[\mathring{\varphi},\mathring{\pi}\right]}{\delta\mathring{\varphi}_{B}\left(y\right)\delta\mathring{\varphi}_{A}\left(x\right)}\nonumber \\
 & -\frac{\delta^{2}H\left[\mathring{\varphi},\mathring{\pi}\right]}{\delta\mathring{\varphi}_{B}\left(y\right)\delta\mathring{\varphi}_{A}\left(x\right)}\frac{\delta^{2}S\left[\mathring{\varphi},\mathring{\pi}\right]}{\delta\mathring{\pi}_{C}\left(z\right)\delta\mathring{\pi}_{A}\left(x\right)}-\frac{\delta^{2}H\left[\mathring{\varphi},\mathring{\pi}\right]}{\delta\mathring{\pi}_{C}\left(z\right)\delta\mathring{\varphi}_{A}\left(x\right)}\frac{\delta^{2}S\left[\mathring{\varphi},\mathring{\pi}\right]}{\delta\mathring{\varphi}_{B}\left(y\right)\delta\mathring{\pi}_{A}\left(x\right)}\Bigg]\Delta\varphi_{B}\left(y\right)\Delta\pi_{C}\left(z\right)\nonumber \\
+ & \Bigg[\frac{\delta^{2}H\left[\mathring{\varphi},\mathring{\pi}\right]}{\delta\mathring{\pi}_{B}\left(y\right)\delta\mathring{\pi}_{A}\left(x\right)}\frac{\delta^{2}S\left[\mathring{\varphi},\mathring{\pi}\right]}{\delta\mathring{\pi}_{C}\left(z\right)\delta\mathring{\varphi}_{A}\left(x\right)}-\frac{\delta^{2}H\left[\mathring{\varphi},\mathring{\pi}\right]}{\delta\mathring{\pi}_{B}\left(y\right)\delta\mathring{\varphi}_{A}\left(x\right)}\frac{\delta^{2}S\left[\mathring{\varphi},\mathring{\pi}\right]}{\delta\mathring{\pi}_{C}\left(z\right)\delta\mathring{\pi}_{A}\left(x\right)}\Bigg]\Delta\pi_{B}\left(y\right)\Delta\pi_{C}\left(z\right)\!\Bigg\}\nonumber \\
+ & \mathcal{O}\left(\Delta^{3}\right).\label{eq:GR_Sdot_quadratic_form_general}
\end{align}
\end{widetext}
We compute this, in turn, for a scalar field in curved spacetime,
for EM in curved spacetime, and for GR. We will show
that no function $S$ obeying the conditions S1-S2 given here exists
in the case of the first two, but that the same cannot be said of
the latter.


\subsubsection{\label{sec:4.2.1}Scalar field}

Let us consider a theory for a scalar field $\phi\left(x\right)$
in a potential $V\left[\phi\left(x\right)\right]$, defined by the
Lagrangian 
\begin{equation}
L=\sqrt{-g}\left(-\frac{1}{2}g^{ab}\nabla_{a}\phi\nabla_{b}\phi-V\left[\phi\right]\right).\label{eq:GR_scalar_Lagrangian}
\end{equation}
There are no constraints in this case. For turning the above [Eq.~\eqref{eq:GR_scalar_Lagrangian}] into a canonical theory, let us
choose a foliation of $\mathscr{M}$ such that $\bm{N}=0$. The canonical
measure \cite{crnkovic_covariant_1989} is then simply given by $\bm{\Omega}=\int_{\Sigma}{\rm d}^{3}x\,\delta\dot{\phi}\wedge\delta\phi$,
and the Hamiltonian \cite{poisson_relativists_2007} is 
\begin{equation}
H\left[\phi,\pi\right]\!=\!\int_{\Sigma}\!{\rm d}^{3}x\, N\!\left(\frac{\pi^{2}}{2\sqrt{h}}+\frac{\sqrt{h}}{2}h^{ab}\nabla_{a}\phi\nabla_{b}\phi+\sqrt{h}V\left[\phi\right]\right)\!,\label{eq:GR_scalar_Hamiltonian}
\end{equation}
where $\pi=(\sqrt{h}/N)\dot{\phi}$ is the canonical momentum.

Let us compute the second functional derivatives of $H$. We have:
\begin{align}
\frac{\delta^{2}H\left[\phi,\pi\right]}{\delta\phi\left(y\right)\delta\phi\left(x\right)}= & N\left(x\right)\sqrt{h\left(x\right)}V''\left[\phi\left(x\right)\right]\delta\left(x-y\right)\nonumber\\
 & -\partial_{a}\left(N\left(x\right)\sqrt{h\left(x\right)}h^{ab}\left(x\right)\partial_{b}\delta\left(x-y\right)\right),\label{eq:GR_scalar_d2H_phi}\\
\frac{\delta^{2}H\left[\phi,\pi\right]}{\delta\pi\left(y\right)\delta\pi\left(x\right)}= & \frac{N\left(x\right)}{\sqrt{h\left(x\right)}}\delta\left(x-y\right),\label{eq:GR_scalar_d2H_pi}
\end{align}
and the mixed derivatives $\delta^{2}H\left[\phi,\pi\right]/\delta\pi\left(y\right)\delta\phi\left(x\right)$
vanish.

We now proceed as outlined above: We assume there exists an entropy
function $S:\mathscr{P}\rightarrow\mathbb{R}$ obeying S1-S2 with
an equilibrium field configuration $(\mathring{\phi},\mathring{\pi})$,
and we will show that there is a contradiction with $\dot{S}>0$.
Additionally, we assume that $V''[\mathring{\phi}]\geq0$; in other
words, the equilibrium field configuration is one where the potential
is concave upwards, i.e. it is a stable equilibrium.

According to the above expression for $\dot{S}$ [Eq.~\eqref{eq:GR_Sdot_quadratic_form_general}],
we have that entropy production in this case is given by 
\begin{widetext}
\begin{align}
\dot{S}=\int_{\Sigma}{\rm d}^{3}x\,{\rm d}^{3}y\,{\rm d}^{3}z\,\Bigg\{ & \Bigg[-\frac{\delta^{2}H[\mathring{\phi},\mathring{\pi}]}{\delta\mathring{\phi}\left(y\right)\delta\mathring{\phi}\left(x\right)}\frac{\delta^{2}S[\mathring{\phi},\mathring{\pi}]}{\delta\mathring{\phi}\left(z\right)\delta\mathring{\pi}\left(x\right)}\Bigg]\Delta\phi\left(y\right)\Delta\phi\left(z\right)\nonumber \\
+ & \Bigg[\frac{\delta^{2}H[\mathring{\phi},\mathring{\pi}]}{\delta\mathring{\pi}\left(z\right)\delta\mathring{\pi}\left(x\right)}\frac{\delta^{2}S[\mathring{\phi},\mathring{\pi}]}{\delta\mathring{\phi}\left(y\right)\delta\mathring{\phi}\left(x\right)}-\frac{\delta^{2}H[\mathring{\phi},\mathring{\pi}]}{\delta\mathring{\phi}\left(y\right)\delta\mathring{\phi}\left(x\right)}\frac{\delta^{2}S[\mathring{\phi},\mathring{\pi}]}{\delta\mathring{\pi}\left(z\right)\delta\mathring{\pi}\left(x\right)}\Bigg]\Delta\phi\left(y\right)\Delta\pi\left(z\right)\nonumber \\
+ & \Bigg[\frac{\delta^{2}H[\mathring{\phi},\mathring{\pi}]}{\delta\mathring{\pi}\left(y\right)\delta\mathring{\pi}\left(x\right)}\frac{\delta^{2}S[\mathring{\phi},\mathring{\pi}]}{\delta\mathring{\pi}\left(z\right)\delta\mathring{\phi}\left(x\right)}\Bigg]\Delta\pi\left(y\right)\Delta\pi\left(z\right)\Bigg\}+\mathcal{O}\left(\Delta^{3}\right),\label{eq:GR_scalar_Sdot_quadratic_form}
\end{align}
\end{widetext}
where we have used the fact that the mixed derivatives vanish. Let
us now evaluate $\dot{S}$ along different directions in $\mathscr{P}$
away from $(\mathring{\phi},\mathring{\pi})$. Suppose $\Delta\pi$
is nonzero everywhere on $\Sigma$, and $\Delta\phi$ vanishes everywhere
on $\Sigma$. Then, using the second momentum derivative of $H$ [Eq.~\eqref{eq:GR_scalar_d2H_pi}], $\dot{S}$ [Eq.~\eqref{eq:GR_scalar_Sdot_quadratic_form}]
becomes: 
\begin{align}
\dot{S}= & \!\int_{\Sigma}\!\!{\rm d}^{3}x\,{\rm d}^{3}y\,{\rm d}^{3}z\,\frac{N\left(x\right)}{\sqrt{h\left(x\right)}}\delta\left(x-y\right)\nonumber\\
 & \times\frac{\delta^{2}S[\mathring{\phi},\mathring{\pi}]}{\delta\mathring{\pi}\left(z\right)\delta\mathring{\phi}\left(x\right)}\Delta\pi\left(y\right)\Delta\pi\left(z\right)+\mathcal{O}\left(\Delta^{3}\right)\label{eq:GR_scalar_Sdot_dpi_1}\\
= & \!\int_{\Sigma}\!\!{\rm d}^{3}y\,{\rm d}^{3}z\,\frac{N\left(y\right)}{\sqrt{h\left(y\right)}}\frac{\delta^{2}S[\mathring{\phi},\mathring{\pi}]}{\delta\mathring{\pi}\left(z\right)\delta\mathring{\phi}\left(y\right)}\Delta\pi\left(y\right)\Delta\pi\left(z\right)\!+\!\mathcal{O}\!\left(\Delta^{3}\right)\label{eq:GR_scalar_Sdot_dpi_2}\\
\leq & \left\{ \max_{x\in\Sigma}\frac{N\left(x\right)}{\sqrt{h\left(x\right)}}\left(\Delta\pi\left(x\right)\right)^{2}\right\} \int_{\Sigma}\!{\rm d}^{3}y\,{\rm d}^{3}z\,\frac{\delta^{2}S[\mathring{\phi},\mathring{\pi}]}{\delta\mathring{\pi}\left(z\right)\delta\mathring{\phi}\left(y\right)}\nonumber\\
 & +\mathcal{O}\left(\Delta^{3}\right).\label{eq:GR_scalar_Sdot_dpi_3}
\end{align}
The requirement that the LHS of the first line above [Eq.~\eqref{eq:GR_scalar_Sdot_dpi_1}]
is strictly positive, combined with the strict positivity of the term
in curly brackets in the third line [Eq.~\eqref{eq:GR_scalar_Sdot_dpi_3}]
and the assumption (S2) of the definiteness of the Hessian of $\dot{S}$
at $(\mathring{\phi},\mathring{\pi})$, altogether mean that the above
[Eqs.~\eqref{eq:GR_scalar_Sdot_dpi_1}-\eqref{eq:GR_scalar_Sdot_dpi_3}]
imply: 
\begin{equation}
\int_{\Sigma}{\rm d}^{3}y\,{\rm d}^{3}z\,\frac{\delta^{2}S[\mathring{\phi},\mathring{\pi}]}{\delta\mathring{\pi}\left(z\right)\delta\mathring{\phi}\left(y\right)}>0.\label{eq:GR_scalar_Spiphi}
\end{equation}
Now let us evaluate $\dot{S}$ in a region of $\mathscr{P}$ where
$\Delta\phi$ is nonzero everywhere on $\Sigma$, while $\Delta\pi$
vanishes everywhere on $\Sigma$. Then, using the second field derivative
of $H$ [Eq.~\eqref{eq:GR_scalar_d2H_phi}], the negative of the
above expression for $\dot{S}$ [Eq.~\eqref{eq:GR_scalar_Sdot_quadratic_form}]
becomes: 
\begin{align}
-\dot{S}= & \int_{\Sigma}{\rm d}^{3}x\,{\rm d}^{3}y\,{\rm d}^{3}z\,\bigg\{ N\left(x\right)\sqrt{h\left(x\right)}V''[\mathring{\phi}\left(x\right)]\delta\left(x-y\right)\nonumber \\
 & -\partial_{a}\left(N\left(x\right)\sqrt{h\left(x\right)}h^{ab}\left(x\right)\partial_{b}\delta\left(x-y\right)\right)\bigg\}\nonumber\\
 & \times\frac{\delta^{2}S[\mathring{\phi},\mathring{\pi}]}{\delta\mathring{\phi}\left(z\right)\delta\mathring{\pi}\left(x\right)}\Delta\phi\left(y\right)\Delta\phi\left(z\right)+\mathcal{O}\left(\Delta^{3}\right).\label{eq:GR_scalar_Sdot_dphi_1}
\end{align}
Now, observe that 
\begin{multline}
\int_{\Sigma}{\rm d}^{3}x\,{\rm d}^{3}y\,{\rm d}^{3}z\,\left\{ \partial_{a}\left(N\left(x\right)\sqrt{h\left(x\right)}h^{ab}\left(x\right)\partial_{b}\delta\left(x-y\right)\right)\right\}\\
\times\frac{\delta^{2}S[\mathring{\phi},\mathring{\pi}]}{\delta\mathring{\phi}\left(z\right)\delta\mathring{\pi}\left(x\right)}\Delta\phi\left(y\right)\Delta\phi\left(z\right)\label{eq:GR_scalar_boundary_term}
\end{multline}
is simply a boundary term. This can be seen by integrating by parts
until the derivative is removed from the delta distribution, the definition
of the latter is applied to remove the $x$ integration, and the result
is a total derivative in the integrand. Assuming asymptotic decay
properties sufficient to make this boundary term vanish, the above
$-\dot{S}$ [Eq.~\eqref{eq:GR_scalar_Sdot_dphi_1}] simply becomes:
\begin{align}
-\dot{S}= & \int_{\Sigma}{\rm d}^{3}y\,{\rm d}^{3}z\, N\left(y\right)\sqrt{h\left(y\right)}V''[\mathring{\phi}\left(y\right)]\nonumber\\
 & \times\frac{\delta^{2}S[\mathring{\phi},\mathring{\pi}]}{\delta\mathring{\phi}\left(z\right)\delta\mathring{\pi}\left(y\right)}\Delta\phi\left(y\right)\Delta\phi\left(z\right)+\mathcal{O}\left(\Delta^{3}\right)\label{eq:GR_scalar_Sdot_dphi_2}\\
\geq & \left\{ \min_{x\in\Sigma}N\left(x\right)\sqrt{h\left(x\right)}V''[\mathring{\phi}\left(x\right)]\left(\Delta\phi\left(x\right)\right)^{2}\right\}\nonumber\\
 & \times\int_{\Sigma}{\rm d}^{3}y\,{\rm d}^{3}z\,\frac{\delta^{2}S[\mathring{\phi},\mathring{\pi}]}{\delta\mathring{\phi}\left(z\right)\delta\mathring{\pi}\left(y\right)}+\mathcal{O}\left(\Delta^{3}\right).\label{eq:GR_scalar_Sdot_dphi_3}
\end{align}
The LHS of the first line [Eq.~\eqref{eq:GR_scalar_Sdot_dphi_2}]
should be strictly negative, and the term in curly brackets in the
second line [Eq.~\eqref{eq:GR_scalar_Sdot_dphi_3}] is strictly
positive. Hence, owing to the definiteness of the Hessian of $\dot{S}$
at $(\mathring{\phi},\mathring{\pi})$, and using the symmetry of
the arguments in the integrand and equality of mixed derivatives,
the above [Eqs.~\eqref{eq:GR_scalar_Sdot_dphi_2}-\eqref{eq:GR_scalar_Sdot_dphi_3}]
imply: 
\begin{equation}
\int_{\Sigma}{\rm d}^{3}y\,{\rm d}^{3}z\,\frac{\delta^{2}S[\mathring{\phi},\mathring{\pi}]}{\delta\mathring{\pi}\left(z\right)\delta\mathring{\phi}\left(y\right)}<0.\label{eq:GR_scalar_Spiphi_contradiction}
\end{equation}
This is a contradiction with the inequality obtained previously [Eq.~\eqref{eq:GR_scalar_Spiphi}]. Therefore, we have no function $S$
for a scalar field theory that behaves like entropy according to assumptions
S1-S2.

We remark that in this case, we get the conclusion $\dot{S}=0$ using
the perturbative approach despite the fact that the topological one would
not work in the case of a non-compact Cauchy surface. The reason is
that:
\begin{align}
\mu\left(\mathscr{P}\right)=\int_{\mathscr{P}}\!\!\bm{\Omega} & =\int_{\mathscr{P}}\int_{\Sigma}{\rm d}^{3}x\,\delta\dot{\phi}\left(x\right)\wedge\delta\phi\left(x\right)\label{eq:GR_scalar_mu1}\\
 & \geq\int_{\mathscr{P}}\int_{\Sigma}{\rm d}^{3}x\,\min_{y\in\Sigma}\left[\delta\dot{\phi}\left(y\right)\wedge\delta\phi\left(y\right)\right]\label{eq:GR_scalar_mu2}\\
 & =\int_{\mathscr{P}}\left\{ \min_{y\in\Sigma}\left[\delta\dot{\phi}\left(y\right)\wedge\delta\phi\left(y\right)\right]\right\} \left[\int_{\Sigma}{\rm d}^{3}x\right]\!,\label{eq:GR_scalar_mu3}
\end{align}
which diverges if $\Sigma$ is non-compact. (N.B. The reason why the
term in curly brackets is finite but non-zero is that the field and
its time derivative cannot be always vanishing at any given point, for if they
were it would lead only to the trivial solution.) Thus, only the perturbative
approach is useful here for deducing lack of entropy production for
spacetimes with non-compact Cauchy surfaces.


\subsubsection{\label{sec:4.2.2}Electromagnetism}

Before we inspect EM in curved spacetime, let us carry out the analysis
in flat spacetime ($N=1$, $\bm{N}=\bm{0}$, and $\bm{h}={}^{(3)}\bm{\delta}={\rm diag(0,1,1,1)}$),
for massive (or de Broglie-Proca) EM \cite{prescod-weinstein_extension_2014}, defined by the Lagrangian 
\begin{equation}
L=-\frac{1}{4}\bm{F}:\bm{F}-\frac{1}{2}m^{2}\bm{A}\cdot\bm{A}+\bm{A}\cdot\bm{J},\label{eq:GR_massive_EM_Lagrangian}
\end{equation}
where $F_{ab}=\partial_{a}A_{b}-\partial_{b}A_{a}$, $A_{a}$ is the
electromagnetic potential, and $J_{a}$ is an external source.

We have a constrained Hamiltonian system in this case. In particular,
the momentum canonically conjugate to $A_{0}=V$ vanishes identically.
This means that instead of $A_{a}$, we may take (its spatial part)
$\mathcal{A}_{a}={}^{(3)}\delta_{ab}A^{b}$ along with its conjugate
momentum, $\pi^{a}=\dot{\mathcal{A}}^{a}-\partial^{a}V$, to be the
phase space variables -- while appending to the canonical equations
of motion resulting from $H\left[\bm{\mathcal{A}},\bm{\pi}\right]$
the constraint $0=\delta H/\delta V$. In particular, we have \cite{prescod-weinstein_extension_2014}:
\begin{align}
H\left[\bm{\mathcal{A}},\bm{\pi}\right]=\!\!\int_{\Sigma}\!\!{\rm d}^{3}x & \bigg(\frac{1}{4}\bm{\mathcal{F}}:\bm{\mathcal{F}}+\!\frac{m^{2}}{2}\left(\bm{\mathcal{A}}\cdot\bm{\mathcal{A}}-V^{2}\right)\!-\bm{\mathcal{A}}\cdot\bm{\mathcal{J}}\nonumber \\
 & +\frac{1}{2}\bm{\pi}\cdot\bm{\pi}-\left(\partial_{a}\pi^{a}+\rho\right)V+\partial_{a}\left(V\pi^{a}\right)\!\bigg)\!,\label{eq:GR_massive_EM_Hamiltonian}
\end{align}
where $\mathcal{F}_{ab}={}^{(3)}\delta_{ac}{}^{(3)}\delta_{bd}F^{cd}$,
$\rho=J^{0}$ and $\mathcal{J}^{a}={}^{(3)}\delta^{ab}J_{b}$.

The Poisson bracket [Eq.~\eqref{eq:GR_Sdot_quadratic_form_general}]
is, in this case: 
\begin{widetext}
\begin{align}
\dot{S}=\int_{\Sigma}{\rm d}^{3}x\,{\rm d}^{3}y\,{\rm d}^{3}z\,\Bigg\{ & \Bigg[-\frac{\delta^{2}H[\mathring{\bm{\mathcal{A}}},\mathring{\bm{\pi}}]}{\delta\mathring{\mathcal{A}}_{b}\left(y\right)\delta\mathring{\mathcal{A}}_{a}\left(x\right)}\frac{\delta^{2}S[\mathring{\bm{\mathcal{A}}},\mathring{\bm{\pi}}]}{\delta\mathring{\mathcal{A}}_{c}\left(z\right)\delta\mathring{\pi}^{a}\left(x\right)}\Bigg]\Delta\mathcal{A}_{b}\left(y\right)\Delta\mathcal{A}_{c}\left(z\right)\nonumber \\
+ & \Bigg[\frac{\delta^{2}H[\mathring{\bm{\mathcal{A}}},\mathring{\bm{\pi}}]}{\delta\mathring{\pi}_{c}\left(z\right)\delta\mathring{\pi}_{a}\left(x\right)}\frac{\delta^{2}S[\mathring{\bm{\mathcal{A}}},\mathring{\bm{\pi}}]}{\delta\mathring{\mathcal{A}}_{b}\left(y\right)\delta\mathring{\mathcal{A}}^{a}\left(x\right)}-\frac{\delta^{2}H[\mathring{\bm{\mathcal{A}}},\mathring{\bm{\pi}}]}{\delta\mathring{\mathcal{A}}_{b}\left(y\right)\delta\mathring{\mathcal{A}}_{a}\left(x\right)}\frac{\delta^{2}S[\mathring{\bm{\mathcal{A}}},\mathring{\bm{\pi}}]}{\delta\mathring{\pi}_{c}\left(z\right)\delta\mathring{\pi}^{a}\left(x\right)}\Bigg]\Delta\mathcal{A}_{b}\left(y\right)\Delta\pi_{c}\left(z\right)\nonumber \\
+ & \Bigg[\frac{\delta^{2}H[\mathring{\bm{\mathcal{A}}},\mathring{\bm{\pi}}]}{\delta\mathring{\pi}_{b}\left(y\right)\delta\mathring{\pi}_{a}\left(x\right)}\frac{\delta^{2}S[\mathring{\bm{\mathcal{A}}},\mathring{\bm{\pi}}]}{\delta\mathring{\pi}_{c}\left(z\right)\delta\mathring{\mathcal{A}}^{a}\left(x\right)}\Bigg]\Delta\pi_{b}\left(y\right)\Delta\pi_{c}\left(z\right)\Bigg\}+\mathcal{O}\left(\Delta^{3}\right),\label{eq:GR_EM_Sdot_quadratic_form}
\end{align}
\end{widetext}
where we have used the fact that the mixed derivatives of the Hamiltonian
[Eq.~\eqref{eq:GR_massive_EM_Hamiltonian}] vanish by inspection,
and we compute the second field and momentum derivatives thereof to
be, respectively:
\begin{align}
\frac{\delta^{2}H[\mathring{\bm{\mathcal{A}}},\mathring{\bm{\pi}}]}{\delta\mathring{\mathcal{A}}_{b}\left(y\right)\delta\mathring{\mathcal{A}}_{a}\left(x\right)}\!= & \!-\left\{\! ^{(3)}\delta^{ab}\partial^{c}\partial_{c}\delta\left(x-y\right)-\partial^{b}\partial^{a}\delta\left(x-y\right)\!\right\}\nonumber\\
 & +m^{2}\left[^{(3)}\delta^{ab}\delta\left(x-y\right)\right],\label{eq:GR_massive_EM_HAA}\\
\frac{\delta^{2}H[\mathring{\bm{\mathcal{A}}},\mathring{\bm{\pi}}]}{\delta\mathring{\pi}_{b}\left(y\right)\delta\mathring{\pi}_{a}\left(x\right)}\!= & ^{(3)}\delta^{ab}\delta\left(x-y\right).\label{eq:GR_massive_EM_Hpipi}
\end{align}
Analogously with our strategy in the scalar field case, let us
evaluate $\dot{S}$ along different directions away from equilibrium.
In particular, let us suppose $\Delta\pi_{1}$ is nonzero everywhere
on $\Sigma$, and that $\Delta\pi_{2}$, $\Delta\pi_{3}$, and $\Delta\mathcal{A}_{a}$
all vanish everywhere on $\Sigma$. Then, using the second momentum
derivative of $H$ [Eq.~\eqref{eq:GR_massive_EM_Hpipi}], $\dot{S}$
[Eq.~\eqref{eq:GR_EM_Sdot_quadratic_form}] becomes:
\begin{align}
\dot{S}= & \!\!\int_{\Sigma}\!\!{\rm d}^{3}x\,{\rm d}^{3}y\,{\rm d}^{3}z\,\delta\left(x-y\right)\frac{\delta^{2}S[\mathring{\bm{\mathcal{A}}},\mathring{\bm{\pi}}]}{\delta\mathring{\pi}_{1}\left(z\right)\delta\mathring{\mathcal{A}}_{1}\!\left(x\right)}\Delta\pi_{1}\!\left(y\right)\Delta\pi_{1}\left(z\right)\nonumber\\
 & +\mathcal{O}\left(\Delta^{3}\right)\label{eq:GR_massive_EM_Sdot_dpi_1}\\
= & \!\!\int_{\Sigma}{\rm d}^{3}y\,{\rm d}^{3}z\,\frac{\delta^{2}S[\mathring{\bm{\mathcal{A}}},\mathring{\bm{\pi}}]}{\delta\mathring{\pi}_{1}\left(z\right)\delta\mathring{\mathcal{A}}_{1}\left(y\right)}\Delta\pi_{1}\left(y\right)\Delta\pi_{1}\left(z\right)+\mathcal{O}\left(\Delta^{3}\right)\label{eq:GR_massive_EM_Sdot_dpi_2}\\
\leq & \!\left\{ \max_{x\in\Sigma}\left(\Delta\pi_{1}\left(x\right)\right)^{2}\right\} \!\int_{\Sigma}\!\!{\rm d}^{3}y\,{\rm d}^{3}z\,\frac{\delta^{2}S[\mathring{\bm{\mathcal{A}}},\mathring{\bm{\pi}}]}{\delta\mathring{\pi}_{1}\left(z\right)\delta\mathring{\mathcal{A}}_{1}\left(y\right)}\!\!+\mathcal{O}\!\left(\Delta^{3}\right)\!.\label{eq:GR_massive_EM_Sdot_dpi_3}
\end{align}
The argument proceeds as before: the strict positivity of the LHS
of the first line above [Eq.~\eqref{eq:GR_massive_EM_Sdot_dpi_1}],
combined with that of the term in curly brackets in the third line
[Eq.~\eqref{eq:GR_massive_EM_Sdot_dpi_3}] and the assumption
(S2) of the definiteness of the Hessian of $\dot{S}$ at $(\mathring{\bm{\mathcal{A}}},\mathring{\bm{\pi}})$,
altogether mean that the above [Eqs.~\eqref{eq:GR_massive_EM_Sdot_dpi_1}-\eqref{eq:GR_massive_EM_Sdot_dpi_3}]
imply: 
\begin{equation}
\int_{\Sigma}{\rm d}^{3}y\,{\rm d}^{3}z\,\frac{\delta^{2}S[\mathring{\bm{\mathcal{A}}},\mathring{\bm{\pi}}]}{\delta\mathring{\pi}_{1}\left(z\right)\delta\mathring{\mathcal{A}}_{1}\left(y\right)}>0.\label{eq:GR_massive_EM_SpiA}
\end{equation}
Now let us evaluate $\dot{S}$ where $\Delta\mathcal{A}_{1}$ is nonzero
everywhere on $\Sigma$, while $\Delta\mathcal{A}_{2}$, $\Delta\mathcal{A}_{3}$
and $\Delta\pi_{a}$ all vanish everywhere on $\Sigma$. Then, using
the second field derivative of $H$ [Eq.~\eqref{eq:GR_massive_EM_HAA}],
the negative of the above expression for $\dot{S}$ [Eq.~\eqref{eq:GR_EM_Sdot_quadratic_form}]
becomes:
\begin{multline}
-\dot{S}=\int_{\Sigma}{\rm d}^{3}x\,{\rm d}^{3}y\,{\rm d}^{3}z\,\Bigg[\Big(-\big\{ ^{(3)}\delta^{ab}\partial^{c}\partial_{c}\delta\left(x-y\right)\\
-\partial^{b}\partial^{a}\delta\left(x-y\right)\big\}+m^{2}\left[^{(3)}\delta^{ab}\delta\left(x-y\right)\right]\Big)\frac{\delta^{2}S[\mathring{\bm{\mathcal{A}}},\mathring{\bm{\pi}}]}{\delta\mathring{\mathcal{A}}_{c}\left(z\right)\delta\mathring{\pi}^{a}\left(x\right)}\Bigg]\\
\times\Delta\mathcal{A}_{b}\left(y\right)\Delta\mathcal{A}_{c}\left(z\right)+\mathcal{O}\left(\Delta^{3}\right).\label{eq:GR_massive_EM_Sdot_dA}
\end{multline}
The term in curly brackets simply furnishes a (vanishing) boundary
term (up to $\mathcal{O}(\Delta^{3})$). Note that for $m=0$ (corresponding
to Maxwellian EM in flat spacetime) we would thus get an indefinite
Hessian of $\dot{S}$ at $(\mathring{\bm{\mathcal{A}}},\mathring{\bm{\pi}})$,
and hence no function $S$ that behaves like entropy as per S1-S2.
So let us assume $m^{2}>0$. Using the symmetry of the arguments in
the integrand and equality of mixed derivatives, we are thus left
with:
\begin{align}
-\dot{S}= & \int_{\Sigma}{\rm d}^{3}y\,{\rm d}^{3}z\, m^{2}\frac{\delta^{2}S[\mathring{\bm{\mathcal{A}}},\mathring{\bm{\pi}}]}{\delta\mathring{\pi}_{1}\left(z\right)\delta\mathring{\mathcal{A}}_{1}\left(y\right)}\Delta\mathcal{A}_{1}\left(y\right)\Delta\mathcal{A}_{1}\left(z\right)\nonumber\\
 & +\mathcal{O}\left(\Delta^{3}\right)\label{eq:GR_massive_EM_Sdot_dA_1}\\
\geq & \left\{ m^{2}\min_{x\in\Sigma}\left(\Delta\mathcal{A}_{1}\left(x\right)\right)^{2}\right\} \int_{\Sigma}{\rm d}^{3}y\,{\rm d}^{3}z\,\frac{\delta^{2}S[\mathring{\bm{\mathcal{A}}},\mathring{\bm{\pi}}]}{\delta\mathring{\pi}_{1}\left(z\right)\delta\mathring{\mathcal{A}}_{1}\left(y\right)}\nonumber\\
 & +\mathcal{O}\left(\Delta^{3}\right).\label{eq:GR_massive_EM_Sdot_dA_2}
\end{align}
The LHS of the first line [Eq.~\eqref{eq:GR_massive_EM_Sdot_dA_1}]
should be strictly negative, and the term in curly brackets in the
second line [Eq.~\eqref{eq:GR_massive_EM_Sdot_dA_2}] is strictly
positive. Hence, owing to the definiteness of the Hessian of $\dot{S}$
at $(\mathring{\bm{\mathcal{A}}},\mathring{\bm{\pi}})$, the above [Eqs.~\eqref{eq:GR_massive_EM_Sdot_dA_1}-\eqref{eq:GR_massive_EM_Sdot_dA_2}]
imply:
\begin{equation}
\int_{\Sigma}{\rm d}^{3}y\,{\rm d}^{3}z\,\frac{\delta^{2}S[\mathring{\bm{\mathcal{A}}},\mathring{\bm{\pi}}]}{\delta\mathring{\pi}_{1}\left(z\right)\delta\mathring{\mathcal{A}}_{1}\left(y\right)}<0.\label{eq:GR_massive_EM_SpiA_2}
\end{equation}
This is a contradiction with the previous inequality on the same quantity
[Eq.~\eqref{eq:GR_massive_EM_SpiA}]. Hence there is no function
$S$ that behaves like entropy (according to S1-S2) for a massive
EM field in flat spacetime.

Let us now carry out the proof for a simple Maxwellian EM field in
curved spacetime, defined by the Lagrangian 
\begin{equation}
L=-\frac{1}{4}\sqrt{-g}\bm{F}:\bm{F},\label{eq:GR_EM_Lagrangian}
\end{equation}
where $F_{ab}=\nabla_{a}A_{b}-\nabla_{b}A_{a}$ and $A_{a}$ is the
electromagnetic potential. As in the scalar field case, we work with
a spacetime foliation such that $\bm{N}=0$.

As with EM in flat spacetime, this is a constrained Hamiltonian system:
the momentum canonically conjugate to $A_{0}=V$ vanishes identically,
meaning again that instead of $A_{a}$, we may take (its spatial part)
$\mathcal{A}_{a}=h_{ab}A^{b}$ along with its conjugate momentum,
$\pi^{a}=(\sqrt{h}/N)h^{ab}(\dot{\mathcal{A}}_{b}-\partial_{b}V)$,
to be the physical phase space variables -- appending to the canonical
equations of motion resulting from $H\left[\bm{\mathcal{A}},\bm{\pi}\right]$
the constraint $0=\delta H/\delta V=\partial_{a}\pi^{a}$ (which is
simply Gauss' law). In particular, we have \cite{prescod-weinstein_extension_2014}:
\begin{equation}
H\left[\bm{\mathcal{A}},\bm{\pi}\right]=\!\!\int_{\Sigma}\!\!{\rm d}^{3}x\left(\frac{1}{4}N\sqrt{h}\,\bm{\mathcal{F}}:\bm{\mathcal{F}}+\frac{N}{2\sqrt{h}}\bm{\pi}\cdot\bm{\pi}+\pi^{a}\partial_{a}V\!\right)\!,\label{eq:GR_EM_Hamiltonian}
\end{equation}
where $\mathcal{F}_{ab}=h_{ac}h_{bd}F^{cd}=D_{a}\mathcal{A}_{b}-D_{b}\mathcal{A}_{a}$.

The Poisson bracket [Eq.~\eqref{eq:GR_Sdot_quadratic_form_general}]
is here given by the same expression as in flat spacetime [Eq.~\eqref{eq:GR_scalar_Sdot_quadratic_form}],
owing to the fact that the mixed derivatives of the Hamiltonian [Eq.~\eqref{eq:GR_EM_Hamiltonian}] vanish. Let us focus on regions in
phase space where $\Delta\bm{\pi}$ vanishes everywhere on $\Sigma,$
but $\Delta\bm{\mathcal{A}}$ is everywhere nonzero. There, 
\begin{multline}
\dot{S}=\int_{\Sigma}{\rm d}^{3}x\,{\rm d}^{3}y\,{\rm d}^{3}z\Bigg[-\frac{\delta^{2}H[\mathring{\bm{\mathcal{A}}},\mathring{\bm{\pi}}]}{\delta\mathring{\mathcal{A}}_{b}\left(y\right)\delta\mathring{\mathcal{A}}_{a}\left(x\right)}\frac{\delta^{2}S[\mathring{\bm{\mathcal{A}}},\mathring{\bm{\pi}}]}{\delta\mathring{\mathcal{A}}_{c}\left(z\right)\delta\mathring{\pi}^{a}\left(x\right)}\Bigg]\\
\times\Delta\mathcal{A}_{b}\left(y\right)\Delta\mathcal{A}_{c}\left(z\right)+\mathcal{O}\left(\Delta^{3}\right).\label{eq:GR_EM_Sdot_dA}
\end{multline}
We compute: 
\begin{align}
\frac{\delta^{2}H[\mathring{\bm{\mathcal{A}}},\mathring{\bm{\pi}}]}{\delta\mathring{\mathcal{A}}_{b}\left(y\right)\delta\mathring{\mathcal{A}}_{a}\left(x\right)}\!= & \!-\!\sqrt{h\left(x\right)}\big\{ \!D^{c}\!\left[N\left(x\right)h^{ab}\left(x\right)D_{c}\delta\left(x-y\right)\right]\nonumber\\
 & -D^{b}\left[N\left(x\right)D^{a}\delta\left(x-y\right)\right]\big\} .\label{eq:GR_EM_d2H_dA}
\end{align}
Inserting this into the above expression for $\dot{S}$ [Eq.~\eqref{eq:GR_EM_Sdot_dA}],
we simply get a (vanishing) boundary term (up to $\mathcal{O}(\Delta^{3})$).
We conclude that we have an indefinite Hessian of $\dot{S}$ at $(\mathring{\bm{\mathcal{A}}},\mathring{\bm{\pi}})$,
and hence no function $S$ that behaves like entropy as per S1-S2.

\subsubsection{\label{sec:4.2.3}Gravity}

The phase space $\mathscr{P}$ of GR is, prior to constraint imposition,
the space of all Riemannian 3-metrics $\bm{h}$ on $\Sigma$, together
with their corresponding conjugate momenta $\bm{\pi}$. The symplectic
form on $\mathscr{P}$, which is also the volume form, is given
by \cite{jezierski_energy_1999}:
\begin{equation}
\bm{\Omega}=\int_{\Sigma}{\rm d}^{3}x\,\delta\pi^{ab}\wedge\delta h_{ab}.\label{eq:GR_symplectic_form}
\end{equation}

However, concordant with its diffeomorphism invariance, GR is a constrained
Hamiltonian system. In particular, the Gauss-Codazzi relations impose
certain restrictions on $\bm{h}$ and $\bm{\pi}$ that determine which
subspaces of $\mathscr{P}$ are dynamically accessible through the
canonical equations of motion (as well as being admissible for initial
conditions). Indeed, diffeomorphism invariance directly implies that
the numerical value of the Hamiltonian should be zero, and therefore
that the Hamiltonian functional should simply be a combination of
these constraints modulo boundary terms.

Let $\bm{r}$ be the outward pointing unit normal on $\partial\Sigma$.
We denote by $\bm{\sigma}$ the induced 2-metric on $\partial\Sigma$,
with $\sigma=\det\left(\bm{\sigma}\right)$, and by $\bm{k}$ its
extrinsic curvature, with $k={\rm tr}\left(\bm{k}\right)$. Then the
gravitational Hamiltonian $H$ is given by \cite{bojowald_canonical_2011,wald_general_1984}: 
\begin{align}
H\left[\bm{h},\bm{\pi}\right]= & \int_{\Sigma}{\rm d}^{3}x\left(NC+\bm{N}\cdot\bm{C}\right)\nonumber\\
 & +\oint_{\partial\Sigma}{\rm d}^{2}x\left(-2N\sqrt{\sigma}k+\frac{2}{N}N_{a}r_{b}\pi^{ab}\right),\label{eq:GR_hamiltonian}
\end{align}
where, using $\pi={\rm tr}\left(\bm{\pi}\right)$, and $\mathcal{R}$
to denote the Ricci scalar on $\Sigma$, 
\begin{align}
C= & -\sqrt{h}\mathcal{R}+\frac{1}{\sqrt{h}}\left(\bm{\pi}:\bm{\pi}-\frac{1}{2}\pi^{2}\right),\label{eq:GR_hamiltonian_C}\\
C_{a}= & -2D^{b}\pi_{ab}.\label{eq:GR_hamiltonian_Ca}
\end{align}
are the Hamiltonian and momentum constraints of GR, respectively.
We can obtain from this Hamiltonian [Eq.~\eqref{eq:GR_hamiltonian}]
the canonical equations of motion $\dot{h}_{ab}=\{h_{ab},H\}=\delta H/\delta\pi^{ab}$
and $\dot{\pi}_{ab}=\{\pi_{ab},H\}=-\delta H/\delta h^{ab}$, in addition
to the constraint equations $C=0=\bm{C}$.

Following the same procedure as before for a hypothetical entropy
functional $S[\bm{h},\bm{\pi}]$ and an equilibrium configuration
$(\mathring{\bm{h}},\mathring{\bm{\pi}})$ in phase space, we see
that the Poisson bracket [Eq.~\eqref{eq:GR_Sdot_quadratic_form_general}]
in this case has the following form: 
\begin{widetext}
\begin{align}
\dot{S}=\int_{\Sigma}{\rm d}^{3}x\,{\rm d}^{3}y\,{\rm d}^{3}z\Bigg\{ & \Bigg[\frac{\delta^{2}H[\mathring{\bm{h}},\mathring{\bm{\pi}}]}{\delta\mathring{h}^{cd}\left(y\right)\delta\mathring{\pi}^{ab}\left(x\right)}\frac{\delta^{2}S[\mathring{\bm{h}},\mathring{\bm{\pi}}]}{\delta\mathring{h}_{ef}\left(z\right)\delta\mathring{h}_{ab}\left(x\right)}-\frac{\delta^{2}H[\mathring{\bm{h}},\mathring{\bm{\pi}}]}{\delta\mathring{h}^{cd}\left(y\right)\delta\mathring{h}^{ab}\left(x\right)}\frac{\delta^{2}S[\mathring{\bm{h}},\mathring{\bm{\pi}}]}{\delta\mathring{h}_{ef}\left(z\right)\delta\mathring{\pi}_{ab}\left(x\right)}\Bigg]\Delta h^{cd}\left(y\right)\Delta h_{ef}\left(z\right)\nonumber \\
+ & \Bigg[\frac{\delta^{2}H[\mathring{\bm{h}},\mathring{\bm{\pi}}]}{\delta\mathring{h}^{cd}\left(y\right)\delta\mathring{\pi}^{ab}\left(x\right)}\frac{\delta^{2}S[\mathring{\bm{h}},\mathring{\bm{\pi}}]}{\delta\mathring{\pi}_{ef}\left(z\right)\delta\mathring{h}_{ab}\left(x\right)}+\frac{\delta^{2}H[\mathring{\bm{h}},\mathring{\bm{\pi}}]}{\delta\mathring{\pi}_{ef}\left(z\right)\delta\mathring{\pi}^{ab}\left(x\right)}\frac{\delta^{2}S[\mathring{\bm{h}},\mathring{\bm{\pi}}]}{\delta\mathring{h}^{cd}\left(y\right)\delta\mathring{h}_{ab}\left(x\right)}\nonumber \\
 & -\frac{\delta^{2}H[\mathring{\bm{h}},\mathring{\bm{\pi}}]}{\delta\mathring{h}^{cd}\left(y\right)\delta\mathring{h}^{ab}\left(x\right)}\frac{\delta^{2}S[\mathring{\bm{h}},\mathring{\bm{\pi}}]}{\delta\mathring{\pi}_{ef}\left(z\right)\delta\mathring{\pi}_{ab}\left(x\right)}-\frac{\delta^{2}H[\mathring{\bm{h}},\mathring{\bm{\pi}}]}{\delta\mathring{\pi}_{ef}\left(z\right)\delta\mathring{h}^{ab}\left(x\right)}\frac{\delta^{2}S[\mathring{\bm{h}},\mathring{\bm{\pi}}]}{\delta\mathring{h}^{cd}\left(y\right)\delta\mathring{\pi}_{ab}\left(x\right)}\Bigg]\Delta h^{cd}\left(y\right)\Delta\pi_{ef}\left(z\right)\nonumber \\
+ & \Bigg[\frac{\delta^{2}H[\mathring{\bm{h}},\mathring{\bm{\pi}}]}{\delta\mathring{\pi}^{cd}\left(y\right)\delta\mathring{\pi}^{ab}\left(x\right)}\frac{\delta^{2}S[\mathring{\bm{h}},\mathring{\bm{\pi}}]}{\delta\mathring{\pi}_{ef}\left(z\right)\delta\mathring{h}_{ab}\left(x\right)}-\frac{\delta^{2}H[\mathring{\bm{h}},\mathring{\bm{\pi}}]}{\delta\mathring{\pi}^{cd}\left(y\right)\delta\mathring{h}^{ab}\left(x\right)}\frac{\delta^{2}S[\mathring{\bm{h}},\mathring{\bm{\pi}}]}{\delta\mathring{\pi}_{ef}\left(z\right)\delta\mathring{\pi}_{ab}\left(x\right)}\Bigg]\Delta\pi^{cd}\left(y\right)\Delta\pi_{ef}\left(z\right)\Bigg\}\nonumber \\
+ & \mathcal{O}\left(\Delta^{3}\right).\label{eq:GR_Sdot_quadratic_form}
\end{align}
\end{widetext}
The difference with the previous cases is that here, in general, none
of the second derivatives of the Hamiltonian vanish, and crucially,
they do not have a definite sign. For example, let us compute the
second derivative of $H$ with respect to the canonical momentum:
\begin{align}
\frac{\delta^{2}H[\mathring{\bm{h}},\mathring{\bm{\pi}}]}{\delta\mathring{\pi}_{cd}\left(y\right)\delta\mathring{\pi}^{ab}\left(x\right)}= & \frac{2\mathring{N}\left(x\right)}{\sqrt{\mathring{h}\left(x\right)}}\left(\delta^{c}{}_{(a}\delta^{d}{}_{b)}-\frac{1}{2}\mathring{h}_{ab}\left(x\right)\delta^{cd}\right)\nonumber\\
 & \times\delta\left(x-y\right).\label{eq:GR_Hpipi}
\end{align}
In CM or the examples of field theories in curved spacetime we have
considered, the second derivative of $H$ with respect to the momentum
had a definite sign (by virtue of its association with the positivity
of kinetic-type terms). In this case, however, this second derivative
[Eq.~\eqref{eq:GR_Hpipi}] is neither always positive nor always
negative. Thus an argument similar to the previous proofs cannot work
here: the gravitational Hamiltonian [Eq.~\eqref{eq:GR_hamiltonian}]
is of such a nature that its concavity in phase space components (as
is, for example, its concavity in the canonical momentum components
[Eq.~\eqref{eq:GR_Hpipi}]) is not independent of the phase space
variables themselves, and cannot be ascribed a definite (positive
or negative) sign. And so, a contradiction cannot arise with the Poisson
bracket of a phase space functional (such as the gravitational entropy)
being non-zero (and, in particular, positive).

\subsection{\label{sec:4.3}Topological approach}

As discussed in subsection \ref{sec:3.3}, the topological proofs
of Olsen for the non-existence of entropy production in CM rely crucially
on the assumption that the phase space $\mathscr{P}$ is compact.
In such a situation, a system has a finite measure of phase space
$\mu(\mathscr{P})$ available to explore, and there cannot exist a
function which continually increases along orbits.

By contrast, in GR, it is believed that the (reduced) phase
space $\mathscr{S}$ is generically noncompact \cite{schiffrin_measure_2012}. That is to say, the
measure $\mu(\mathscr{S})=\int_{\mathscr{S}}\bm{\Omega}|_{\mathscr{S}}$
in general diverges, where $\bm{\Omega}|_{\mathscr{S}}$ is (using
the notation of section \ref{sec:Setup}) the pullback of the symplectic
form [Eq.~\eqref{eq:GR_symplectic_form}] to $\mathscr{S}$. This
means that the same methods of proof as in CM (subsection \ref{sec:3.3})
cannot be applied.

The connection between a (monotonically increasing) entropy function
in GR and the divergence of its (reduced) phase space measure warrants
some discussion. The latter, it may be noted, is arguably not completely
inevitable. In other words, one may well imagine a space of admissible
solutions to the Einstein equations (or equivalently, the canonical
gravitational equations) whose effective degrees of freedom are such
that they form a finite-measure phase space. Dynamically-trivial examples
of this might be SD black holes. Thus the assertion that
$\mu(\mathscr{S})$ diverges hinges on the nature of the degrees of
freedom believed to be available in the spacetimes under consideration.
However, it has been explicitly shown \cite{schiffrin_measure_2012} that even in very basic dynamically-nontrivial
situations, such as simple cosmological spacetimes, $\mu(\mathscr{S})$
does indeed diverge. In fact, the proof found in \cite{schiffrin_measure_2012} is carried out for compact Cauchy surfaces, and the conclusion is therefore in concordance with the no-return theorem \cite{tipler_general_1979,tipler_general_1980} which also assumes compact Cauchy surfaces. In the following section, we will show that this
happens for perturbed SD spacetimes as well (where the Cauchy surface is non-compact).

The generic divergence of $\mu(\mathscr{S})$ entails that a gravitational
system has an unbounded region of phase space available to explore.
In other words, it is not confined to a finite region where it would
have to eventually return to a configuration from which it started
(which would make a monotonically increasing entropy function impossible). 

It is moreover worth remarking that this situation creates nontrivial problems
for a statistical (i.e. probability-based) general-relativistic definition
of entropy, $S:\mathscr{T}\rightarrow\mathbb{R}$ (as described in
Sections \ref{sec:Intro} and \ref{sec:Setup}) -- which, indeed, one may also ultimately desire
to work with and relate to the mechanical meaning of entropy mainly
discussed in this paper. Naively, one might think of defining such
a statistical entropy function as something along the lines of $S=-\sum_{X}P\left(X\right)\ln P\left(X\right)$,
where $X$ denotes a physical property of interest and $P\left(X\right)$
its probability. In turn, the latter might be understood as the relative
size of the phase space region $\mathscr{S}_{X}\subset\mathscr{S}$
possessing the property $X$, i.e. $P\left(X\right)=\mu(\mathscr{S}_{X})/\mu(\mathscr{S})$.
In this case, we either have \cite{schiffrin_measure_2012}: $P\left(X\right)=0$ if $\mu(\mathscr{S}_{X})$
is finite, $P\left(X\right)=1$ if $\mu(\mathscr{S}\backslash\mathscr{S}_{X})$
is finite, or $P\left(X\right)$ is ill-defined otherwise. Ostensibly,
one would need to invoke a regularisation procedure in order to obtain
finite probabilities (in general) according to this. However, different
regularisation procedures that have been applied (mainly in the context
of cosmology) have proven to yield widely different results depending
on the method of the procedure being used \cite{schiffrin_measure_2012}. Alternatively, a statistical
general-relativistic definition of $S$ in terms of a probability
density $\rho:\mathscr{S}\times\mathscr{T}\rightarrow[0,1]$ (similarly
to CM) as $S=-\int_{\mathscr{S}}\bm{\Omega}|_{\mathscr{S}}\rho\ln\rho$
would likewise face divergence issues. Therefore, any future attempt
to define gravitational entropy in such a context will have to either
devise an unambiguous and well-defined regularisation procedure (for
obtaining finite probabilities), or implement a well-justified cutoff
of the (reduced) phase space measure.

We now turn to discussing these issues in a context where we expect
an intuitive illustration of gravitational entropy production -- the
two-body problem.

\section{\label{sec:2BP}Entropy in the gravitational two-body problem}

One of the most elementary situations in GR in which we expect the
manifestation of a phenomenon such as entropy production is the gravitational
two-body problem.

In CM, the two-body (or Kepler) problem manifestly involves no increase
in the entropy of a system. The perturbative approach, as discussed
in subsubsection \ref{sec:3.2.3}, involves assumptions on the nature
of the Hamiltonian which preclude any conclusions from it in this
regard. However, the topological approach, elaborated in subsection
\ref{sec:3.3}, is applicable: assuming that Keplerian orbits are
bounded, the configuration space $\mathscr{Q}$ can be considered
to be compact, and therefore the phase space $\mathscr{P}$ obtained
from it (involving finite conjugate momenta) is compact as well. Concordant
with the topological proofs, then, we will have no entropy production
in such a situation. The case of the $N$-body problem however is, as alluded to earlier, not the same: neither the assumptions of the perturbative approach, not of the topological approach (specifically, a compact phase space) are applicable, and it has been shown that a monotonically increasing function on phase space does in fact exist \cite{barbour_gravitational_2013,barbour_identification_2014}, and hence, a gravitational arrow of time (and entropy production) associated with it.

In GR, we know the two-body problem involves energy loss and therefore
should implicate an associated production of entropy. The no-return theorem \cite{tipler_general_1979,tipler_general_1980} is inapplicable here because this problem does not involve a compact Cauchy surface. The perturbative
approach here fails to disprove the second law (as discussed in subsubsection
\ref{sec:4.2.3}), and we will now show that so too does the topological
approach.

The two-body problem in GR where one small body orbits a much larger
body of mass $M$ can be modeled in the context of perturbations
to the SD metric, 
\begin{equation}
g_{ab}{\rm d}x^{a}{\rm d}x^{b}=-v\left(r\right){\rm d}t^{2}+\frac{{\rm d}r^{2}}{v\left(r\right)}+r^{2}\sigma_{ab}{\rm d}x^{a}{\rm d}x^{b},\label{eq:2BP_Schwarzschild}
\end{equation}
where $v\left(r\right)=1-2M/r$ and $\bm{\sigma}={\rm diag}(0,0,1,\sin^{2}\theta)$ is the metric
of the two-sphere $\mathbb{S}^{2}$. According to standard black hole
perturbation theory (see, for example, \cite{chandrasekhar_mathematical_1998,frolov_black_1998,price_developments_2007}), it is possible to choose
a gauge so that the polar and axial parts of perturbations to this
metric are encoded in a single gauge-invariant variable each. In particular,
they are given respectively by 
\begin{equation}
\Phi_{\left(\pm\right)}=\frac{1}{r}\sum_{l=0}^{\infty}\sum_{m=-l}^{l}Y^{lm}\left(\theta,\phi\right)\Psi_{\left(\pm\right)}^{lm}\left(t,r\right),\label{eq:2BP_perturbations}
\end{equation}
where $Y^{lm}$ are spherical harmonics and $\Psi_{(\pm)}^{lm}$ are
called, respectively, the Zerilli and Regge-Wheeler master functions,
which satisfy known wave-like equations and from which the perturbations
to $\bm{g}$ can be reconstructed. In \cite{jezierski_energy_1999}, the symplectic form of the
reduced phase space $\mathscr{S}$ for such spacetimes is computed.
Without entering into the details of the computation we simply state
the result: 
\begin{equation}
\bm{\Omega}|_{\mathscr{S}}=\sum_{\varsigma=\pm}\int_{\Sigma}{\rm d}^{3}x\,\delta\Upsilon_{\left(\varsigma\right)}\wedge\mathbb{D}\delta\Phi_{\left(\varsigma\right)},\label{eq:2BP_symplectic_form}
\end{equation}
where $\Upsilon_{(\pm)}=[r^{2}\sin\theta/v(r)]\dot{\Phi}_{(\pm)}$,
and $\mathbb{D}=\Delta_{\bm{\sigma}}^{-1}(\Delta_{\bm{\sigma}}+2)^{-1}$
where $\Delta_{\bm{\sigma}}$ is the Laplace operator on $\mathbb{S}^{2}$.

The work \cite{jezierski_energy_1999} where this symplectic form [Eq.~\eqref{eq:2BP_symplectic_form}]
was derived simply uses it to define and formulate conservation laws
for energy and angular momentum in perturbed SD spacetimes.
It does not, however, address the question of the total measure of
$\mathscr{S}$. We will now show that the (reduced) phase space measure
$\mu\left(\mathscr{S}\right)=\int_{\mathscr{S}}\bm{\Omega}|_{\mathscr{S}}$
for such spacetimes in fact diverges, preventing any argument based
on phase space compactness for the non-existence of entropy production.

Inserting the definitions of the different variables and suppressing
for the moment the coordinate dependence of the spherical harmonics
and master functions, we have 
\begin{align}
\mu\left(\mathscr{S}\right)=\, & \sum_{\varsigma=\pm}\int_{\mathscr{S}}\int_{\Sigma}{\rm d}^{3}x\,\delta\Upsilon_{\left(\varsigma\right)}\wedge\mathbb{D}\delta\Phi_{\left(\varsigma\right)}\label{eq:2BP_measure_1}\\
=\, & \sum_{\varsigma=\pm}\int_{\mathscr{S}}\int_{\Sigma}{\rm d}^{3}x\,\delta\left(\frac{r^{2}\sin\theta}{v\left(r\right)}\dot{\Phi}_{\left(\varsigma\right)}\right)\wedge\mathbb{D}\delta\Phi_{\left(\varsigma\right)}\label{eq:2BP_measure_2}\\
=\, & \sum_{\varsigma=\pm}\int_{\mathscr{S}}\int_{\Sigma}{\rm d}^{3}x\,\delta\left(\frac{r\sin\theta}{v\left(r\right)}\sum_{l,m}Y^{lm}\dot{\Psi}_{\left(\varsigma\right)}^{lm}\right)\nonumber\\
 & \quad\quad\quad\quad\quad\wedge\mathbb{D}\delta\left(\frac{1}{r}\sum_{l',m'}Y^{l'm'}\Psi_{\left(\varsigma\right)}^{l'm'}\right).\label{eq:2BP_measure_3}
\end{align}
Now using the fact that the functional exterior derivative acts only
on the master functions and the operator $\mathbb{D}$ only on the
spherical harmonics, we can write this as 
\begin{align}
\mu\left(\mathscr{S}\right)=\, & \sum_{\varsigma=\pm}\int_{\mathscr{S}}\int_{\Sigma}{\rm d}^{3}x\,\left(\frac{r\sin\theta}{v\left(r\right)}\sum_{l,m}Y^{lm}\delta\dot{\Psi}_{\left(\varsigma\right)}^{lm}\right)\nonumber\\
 & \quad\quad\quad\quad\wedge\left(\frac{1}{r}\sum_{l',m'}\left(\mathbb{D}Y^{l'm'}\right)\delta\Psi_{\left(\varsigma\right)}^{l'm'}\right)\label{eq:2BP_measure_4}\\
=\, & \sum_{\varsigma=\pm}\sum_{l,l',m,m'}\int_{\mathscr{S}}\int_{\Sigma}{\rm d}^{3}x\,\left(\frac{r\sin\theta}{v\left(r\right)}Y^{lm}\frac{1}{r}\mathbb{D}Y^{l'm'}\right)\nonumber\\
 & \quad\quad\quad\quad\quad\quad\quad\quad\quad\times\delta\dot{\Psi}_{\left(\varsigma\right)}^{lm}\wedge\delta\Psi_{\left(\varsigma\right)}^{l'm'}.\label{eq:2BP_measure_5}
\end{align}
Writing the Cauchy surface integral in terms of coordinates and collecting
terms, 
\begin{widetext}
\begin{align}
\mu\left(\mathscr{S}\right)=\, & \sum_{\varsigma=\pm}\sum_{l,l',m,m'}\int_{\mathscr{S}}\int_{2M}^{\infty}{\rm d}r\int_{\mathbb{S}^{2}}{\rm d}\theta{\rm d}\phi\,\frac{1}{v\left(r\right)}\left[\left(\sin\theta\right)Y^{lm}\mathbb{D}Y^{l'm'}\right]\delta\dot{\Psi}_{\left(\varsigma\right)}^{lm}\wedge\delta\Psi_{\left(\varsigma\right)}^{l'm'}\label{eq:2BP_measure_6}\\
=\, & \sum_{\varsigma=\pm}\sum_{l,l',m,m'}\int_{\mathscr{S}}\left[\int_{\mathbb{S}^{2}}{\rm d}\theta{\rm d}\phi\,\left(\sin\theta\right)Y^{lm}\mathbb{D}Y^{l'm'}\right]\int_{2M}^{\infty}\frac{{\rm d}r}{v\left(r\right)}\delta\dot{\Psi}_{\left(\varsigma\right)}^{lm}\wedge\delta\Psi_{\left(\varsigma\right)}^{l'm'}\label{eq:2BP_measure_7}\\
=\, & \sum_{\varsigma=\pm}\sum_{l,l',m,m'}A^{ll'mm'}\int_{\mathscr{S}}\int_{2M}^{\infty}\frac{{\rm d}r}{v\left(r\right)}\delta\dot{\Psi}_{\left(\varsigma\right)}^{lm}\wedge\delta\Psi_{\left(\varsigma\right)}^{l'm'},\label{eq:2BP_measure_8}
\end{align}
\end{widetext}
where $A^{ll'mm'}=\int_{\mathbb{S}^{2}}{\rm d}\theta{\rm d}\phi\,(\sin\theta)Y^{lm}\mathbb{D}Y^{l'm'}$
is a finite integral involving only the spherical harmonics. Restoring
the arguments of the master functions, and recalling that the meaning
of $\delta f(t,r)$ (for any function $f$) is simply that a one-form
on the phase space at $(t,r)$ in spacetime, we can write from the
above [Eq.~\eqref{eq:2BP_measure_8}]: 
\begin{widetext}
\begin{align}
\mu\left(\mathscr{S}\right)=\, & \sum_{\varsigma=\pm}\sum_{l,l',m,m'}A^{ll'mm'}\int_{\mathscr{S}}\int_{2M}^{\infty}\frac{{\rm d}r}{v\left(r\right)}\left[\delta\dot{\Psi}_{\left(\varsigma\right)}^{lm}\left(t,r\right)\wedge\delta\Psi_{\left(\varsigma\right)}^{l'm'}\left(t,r\right)\right]\label{eq:2BP_measure_9}\\
\geq\, & \sum_{\varsigma=\pm}\sum_{l,l',m,m'}A^{ll'mm'}\int_{\mathscr{S}}\int_{2M}^{\infty}\frac{{\rm d}r}{v\left(r\right)}\left[\min_{\bar{r}\in[2M,\infty)}\delta\dot{\Psi}_{\left(\varsigma\right)}^{lm}\left(t,\bar{r}\right)\wedge\delta\Psi_{\left(\varsigma\right)}^{l'm'}\left(t,\bar{r}\right)\right]\label{eq:2BP_measure_10}\\
=\, & \sum_{\varsigma=\pm}\sum_{l,l',m,m'}A^{ll'mm'}\left[\int_{\mathscr{S}}\min_{\bar{r}\in[2M,\infty)}\delta\dot{\Psi}_{\left(\varsigma\right)}^{lm}\left(t,\bar{r}\right)\wedge\delta\Psi_{\left(\varsigma\right)}^{l'm'}\left(t,\bar{r}\right)\right]\int_{2M}^{\infty}\frac{{\rm d}r}{v\left(r\right)}\label{eq:2BP_measure_11}\\
=\, & \left\{ \sum_{\varsigma=\pm}\sum_{l,l',m,m'}A^{ll'mm'}\left[\min_{\bar{r}\in[2M,\infty)}\int_{\mathscr{S}}\delta\dot{\Psi}_{\left(\varsigma\right)}^{lm}\left(t,\bar{r}\right)\wedge\delta\Psi_{\left(\varsigma\right)}^{l'm'}\left(t,\bar{r}\right)\right]\right\} \left[\int_{2M}^{\infty}\frac{{\rm d}r}{v\left(r\right)}\right].\label{eq:2BP_measure_12}
\end{align}
\end{widetext}
The phase space integral $\int_{\mathscr{S}}\delta\dot{\Psi}_{(\varsigma)}^{lm}(t,\bar{r})\wedge\delta\Psi_{(\varsigma)}^{l'm'}(t,\bar{r})$
is finite but nonzero even when minimised over $\bar{r}$, because
for any nontrivial solutions of the master functions, there will be
no point in spacetime where they will always be vanishing (for all
time). Thus (assuming that the $l,l',m,m'$ sums are convergent),
everything in the curly bracket in the last line above [Eq.~\eqref{eq:2BP_measure_12}]
is nonzero but finite. However, it multiplies $\int_{2M}^{\infty}{\rm d}r/v(r)$
which diverges (at both integration limits). Hence, $\mu(\mathscr{S})$
diverges for such spacetimes.

We can make a few remarks. Firstly, one might be concerned in the above argument, specifically in the
last line [Eq.~\eqref{eq:2BP_measure_12}], about what might happen
in the asymptotic limit of the phase space integral: in other words,
it maybe the case (i) that $\min_{\bar{r}\in[2M,\infty)}\int_{\mathscr{S}}\delta\dot{\Psi}_{(\varsigma)}^{lm}(t,\bar{r})\wedge\delta\Psi_{(\varsigma)}^{l'm'}(t,\bar{r})$
could turn out to be $\lim_{r\rightarrow\infty}\int_{\mathscr{S}}\delta\dot{\Psi}_{(\varsigma)}^{lm}(t,r)\wedge\delta\Psi_{(\varsigma)}^{l'm'}(t,r)$;
and, if so, one might naively worry (ii) that the latter vanishes
due to asymptotic decay properties of the master functions. This will
actually not happen. To see why, suppose (i) is true. The master functions
must obey outgoing boundary conditions at spatial infinity, i.e. $0=[\partial_{t}+v(r)\partial_{r}]\Psi_{(\varsigma)}^{lm}$
as $r\rightarrow\infty$. Hence we have
\begin{align}
 & \lim_{r\rightarrow\infty}\int_{\mathscr{S}}\delta\dot{\Psi}_{\left(\varsigma\right)}^{lm}\wedge\delta\Psi_{\left(\varsigma\right)}^{l'm'}\nonumber\\
= & \lim_{r\rightarrow\infty}\int_{\mathscr{S}}\delta\left(-v\partial_{r}\Psi_{\left(\varsigma\right)}^{lm}\right)\wedge\delta\Psi_{\left(\varsigma\right)}^{l'm'}\nonumber\\
= & -\int_{\mathscr{S}}\lim_{r\rightarrow\infty}\delta\left(\partial_{r}\Psi_{\left(\varsigma\right)}^{lm}\right)\wedge\delta\Psi_{\left(\varsigma\right)}^{l'm'},\label{eq:2BP_lim}
\end{align}
which is nonzero, because the vanishing of the master functions and
their radial partials at spatial infinity for all time corresponds
only to trivial solutions. Therefore, we have that $\min_{\bar{r}\in[2M,\infty)}\int_{\mathscr{S}}\delta\dot{\Psi}_{(\varsigma)}^{lm}(t,\bar{r})\wedge\delta\Psi_{(\varsigma)}^{l'm'}(t,\bar{r})$
is always nonzero for nontrivial solutions.

Secondly, if the two-body system in this framework is an extreme-mass-ratio
inspiral \cite{blanchet_mass_2011} i.e. the mass of the orbiting body, or ``particle'', is
orders of magnitude smaller than that of the larger one, and the former
is modeled using a stress-energy-momentum tensor with support only
on its worldline, then it is known that $\Psi_{(\pm)}^{lm}(t,r)$
has a discontinuity at the particle location, and thus, $\dot{\Psi}_{(\pm)}^{lm}(t,r)$
has a divergence there. Hence, the integral over $\mathscr{S}$ even
before our inequality above [Eq.~\eqref{eq:2BP_measure_9}] is
already divergent due to the divergence of $\dot{\Psi}_{(\pm)}^{lm}(t,r)$
in the integrand. However, given that such an approach to describing
these systems (i.e. having a stress-energy-momentum tensor of the
particle with a delta distribution) is only an idealization, we regard
the conclusion that $\mu(\mathscr{S})$ diverges as more convincing based on
our earlier argument, which is valid in general -- that is, even for possible descriptions of the smaller body that may be more realistic than that using delta distributions.

\section{\label{sec:Conclusions}Conclusions}

We have proven that there does not exist a monotonically increasing function of phase space -- which may be identified as (what we have referred to as a ``mechanical'' notion of) entropy -- in classical mechanics with $N$ degrees of freedom for certain classes of Hamiltonians, as well as in some (classical) matter field theories in curved (nondynamical) spacetime, \textit{viz.} for standard scalar and electromagnetic fields. To do this, we have followed the procedure for the proof sketched by Poincar\'{e} \cite{poincare_sur_1889} (what we have dubbed the \textsl{perturbative} approach), and we have here carried it out in full rigour for classical mechanics and extended it via similar techniques to field theories. What is noteworthy about this perturbative proof -- counter (to our knowledge) to all other well-known proofs for the non-existence of entropy (in the ``mechanical'' sense) in classical canonical theories -- is that it assumes nothing about the topology of the phase space; in other words, the phase space can be \textsl{non-compact}. Essentially, it relies only on curvature properties (in phase space) of the Hamiltonian of the canonical theory being considered. We have explicated these properties in the case of classical mechanics, and have assumed standard ones for the particular (curved spacetime) matter field theories we have investigated. It would be of interest for future work to determine, in the case of the former, whether they can be made less restrictive (than what we have required for our proof, which thus omits some classes of Hamiltonians of interest such as that for the the gravitational two-body problem), and in the case of the latter, whether they can be generalized or extended to broader classes of field theories. Indeed, it would be in general an interesting question to determine not only the necessary but also -- \textsl{if possible} -- the sufficient conditions that a Hamiltonian of a generic canonical theory needs to satisfy in order for this theorem to be applicable, i.e. in order to preclude ``mechanical'' entropy production. We have seen that it is precisely the curvature properties of the vacuum Hamiltonian of general relativity that prevent this method of proof from being extended thereto, where in fact one does expect (some version of) the second law of thermodynamics to hold.

Topological properties of the phase space can also entail the non-existence of ``mechanical'' entropy, as per the more standard and already well-understood proofs in classical mechanics where the phase space is assumed to be compact \cite{hp1890am,olsen_classical_1993}. However, even for non-gravitational canonical theories this assumption might be too restrictive, and for general relativity, it is believed that in general it is not the case. This renders any of these topological proofs inapplicable in the case of the latter, and moreover, it also significantly complicates any attempt to formulate a sensible ``statistical'' notion of (gravitational) entropy due to the concordant problems in working with finite probabilities (of phase space properties). These must ultimately be overcome (via some regularisation procedure or cutoff argument) for establishing a connection between a ``statistical'' and ``mechanical'' entropy in general relativity. While we still lack any consensus on how to define the latter, it may be hoped that in the future, the generic validity of a (general relativistic) second law may be demonstrated on the basis of (perhaps curvature related) properties of the gravitational Hamiltonian -- which in turn may enter into a statistical mechanics type definition of gravitational entropy in terms of some suitably defined partition function. In this regard, older work based on field-theoretic approaches \cite{horowitz} and more recent developments such as proposals to relate entropy with a Noether charge (specifically, the Noether invariant associated with an infinitesimal time translation) in classical mechanics \cite{Sasa} may provide fruitful hints.

A clear situation in which we anticipate entropy production in general relativity, unlike in classical mechanics, is the gravitational two-body problem. For the latter, as we have discussed, the $N$-body problem actually does also exhibit features of entropy production. We have here shown explicitly that the phase space of perturbed Schwarzschild-Droste spacetimes is non-compact (even without the assumption of self-force). This means that the topological proofs are here inapplicable (but also, on the other hand, so is the ``no-return'' theorem for compact Cauchy surfaces, which by itself cannot be used in this case to understand the non-recurrence of phase space orbits). It is hoped that once a generally agreed upon definition of gravitational entropy is established, one would not only be able to use it to compute the entropy of two-body systems, but also to demonstrate that it should obey the second law (i.e. that it should be monotonically increasing in time). In the long run, an interesting problem to investigate is whether an entropy change, once defined and associated to motion in a Lagrangian formulation, could determine the trajectory of a massive and radiating body, moving in a gravitational field.

\section*{Acknowledgements}

MO would like to thank Richard J. Epp (Waterloo) and C\'{e}sar Ram\'{i}rez Iba\~{n}ez (Waterloo) for useful discussions. MO has received financial support through an Eiffel Bourse d'Excellence n. 840856D by Campus France, and through the Natural Sciences and Engineering Research Council of Canada. MO, LB and ADAMS acknowledge LISA CNES funding. CFS acknowledges support from contracts ESP2013-47637-P and ESP2015-67234-P (Spanish Ministry of Economy and Competitivity of Spain, MINECO). 


\renewcommand{\thesection}{A}
\section{Some details on constrained Hamiltonian systems}\label{sec:A}

In this appendix, we elaborate on some technical details omitted from sections \ref{sec:Setup} and \ref{sec:CM}.

In $\mathscr{P}$, the symplectic form $\bm{\omega}$ is used to define the Hamiltonian
vector field $\bm{X}_{F}$ of any phase space function $F:\mathscr{P}\rightarrow\mathbb{R}$
via $\imath_{\bm{X}_{F}}\bm{\omega}=-{\rm d}F$, where $\imath$ is
the interior product and ${\rm d}$ is the exterior derivative on
$\mathscr{P}$. The Hamiltonian flow (generated by $\bm{X}_{H}$), is determined via ${\rm d}\Phi_{t}/{\rm d}t=\bm{X}_{H}\circ\Phi_{t}$.

\begin{figure}
\begin{centering}
\includegraphics[scale=0.35]{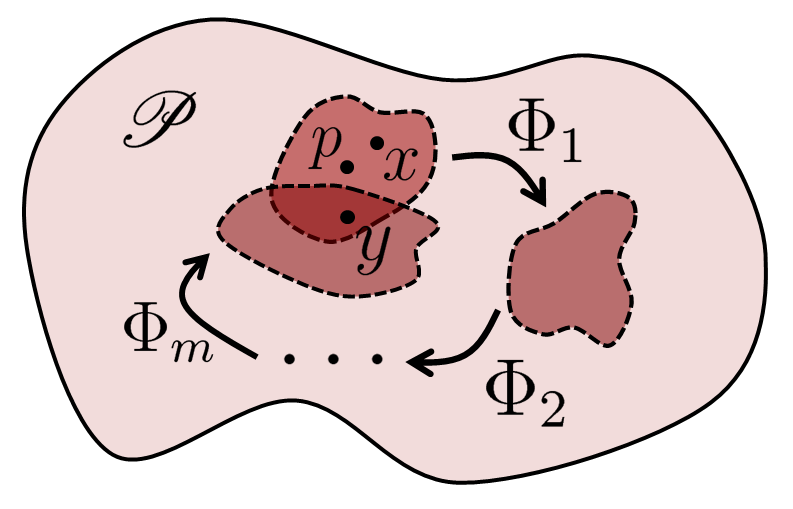}
\par\end{centering}

\protect\caption{\label{fig:recurrence}The idea of the proof for the Poincar\'{e} recurrence theorem.}

\end{figure}

Using this, we can easily offer a proof, shown pictorially in Figure \ref{fig:recurrence}, for the recurrence theorem (see section 16 of \cite{arnold_mathematical_1997}): Assume $\mathscr{P}$ is compact and $\Phi_{t}(\mathscr{P})=\mathscr{P}$.
Let $\mathscr{U}\subset\mathscr{P}$ be the neighborhood of any point
$p\in\mathscr{P}$, and consider the sequence of images $\{\Phi_{n}(\mathscr{U})\}_{n=0}^{\infty}$.
Each $\Phi_{n}(\mathscr{U})$ has the same measure $\int_{\Phi_{n}(\mathscr{U})}\bm{\Omega}$ (because of Liouville's theorem),
so if they never intersected, $\mathscr{P}$ would have infinite measure.
Therefore there exist $k$, $l$ with $k>l$ such that $\Phi_{k}(\mathscr{U})\cap\Phi_{l}(\mathscr{U})\neq\varnothing$,
implying $\Phi_{m}(\mathscr{U})\cap\mathscr{U}\neq\varnothing$ where
$m=k-l$. For any $y\in\Phi_{m}(\mathscr{U})\cap\mathscr{U}$, there
exists an $x\in\mathscr{U}$ such that $y=\Phi_{m}(x)$. Thus, any
point returns arbitrarily close to the initial conditions in a compact and invariant phase space. 

Let us now describe how one must deal with constraints in a Hamiltonian system, i.e. how to go from $(\mathscr{P},\bm{\omega})$ to $(\mathscr{S},\bm{\omega}|_{\mathscr{S}})$ (in the notation already anticipated in section \ref{sec:Setup}). Let $C_{I}:\mathscr{P}\rightarrow\mathbb{R}$ denote a set of phase
space functions indexed by $I$, and suppose the system is subject
to the constraint equations $C_{I}=0$. Then one is only interested
in the constraint surface of the phase space, $\mathscr{C}\subset\mathscr{P}$,
where the constraint equations are satisfied. Let $\bm{\omega}|_{\mathscr{C}}$
denote the pullback of the phase space symplectic form $\bm{\omega}$, onto
the constraint surface $\mathscr{C}$. Depending on the nature of the constraints
$C_{I}$, $\bm{\omega}|_{\mathscr{C}}$ may be degenerate, and therefore
\textsl{not} a symplectic form on $\mathscr{C}$. In particular, if any one
of the constraints is first-class -- meaning that its Hamiltonian vector
field is everywhere tangent to $\mathscr{C}$ -- then $\bm{\omega}|_{\mathscr{C}}$
is degenerate, and hence cannot furnish a symplectic structure. Only
if all constraints are second-class -- i.e. their Hamiltonian vector fields
are nowhere tangent to $\mathscr{C}$ -- will one obtain a pullback
which is itself a symplectic form.

In many situations of interest for classical field theories (as is
the case, for example, in both Maxwellian EM and GR), the constraints
are first-class, and the degeneracy directions of $\bm{\omega}|_{\mathscr{C}}$
correspond precisely to pure gauge variations of the fields. In other
words, the kernel of $\bm{\omega}|_{\mathscr{C}}$ comprises the vector
fields whose flow in phase space represent gauge transformations (of
the $\mathbb{U}\left(1\right)$ gauge symmetry in Maxwellian EM, and
of the diffeomorphism invariance of GR). In principle, a symplectic
form \textsl{could} be obtained if one factors out these vector fields,
by identifying all points on the orbits of their flow. Thus, one could
work with a factor space $\tilde{\mathscr{P}}\subset\mathscr{C}$
which is simply the space of gauge orbits in $\mathscr{C}$, and which
is symplectic.

However, depending on the desired aim of implementing the canonical
construction, taking such an approach can be problematic. This is
because in any theory which is diffeomorphism invariant (such as GR),
``time evolution'' is effected via spacetime diffeomorphisms and
so moving to the space of gauge orbits essentially renders the dynamics
nonexistent: it become entirely trivial, because it is essentially
factored out of $\tilde{\mathscr{P}}$, leaving one with no more sense
of ``motion through phase space''.

There exist two possible solutions for ameliorating this difficulty --
that is, for obtaining a symplectic structure out of $\bm{\omega}|_{\mathscr{C}}$
which \textsl{does} still preserve a nontrivial notion of ``time evolution'':
(a) Instead of passing to the space of gauge orbits, one may instead
choose a representative of each gauge orbit \cite{schiffrin_measure_2012}. The idea is that one
can find a surface $\mathscr{S}\subset\mathscr{C}$ such that each
gauge orbit in $\mathscr{C}$ intersects $\mathscr{S}$ once and only
once. (In fact, sometimes a family of such surfaces that work in localised
regions of $\mathscr{C}$ is needed, but we keep our discussion here
simplified.) The choice of $\mathscr{S}$ is
not unique, and so taking a different surface $\mathscr{S}'$ basically
amounts to a change of description -- corresponding to ``time evolution''
(i.e. change of representative Cauchy surface in spacetime) along
with the associated spatial diffeomorphisms. The technicalities of
this procedure are elaborated in \cite{schiffrin_measure_2012}, but the point is that the subspace
$\mathscr{S}$ of the constraint surface $\mathscr{C}$ resulting
from such a construction \textsl{is} symplectic, whence one can work
with the symplectic form $\bm{\omega}|_{\mathscr{S}}$ obtained by
pulling back $\bm{\omega}|_{\mathscr{C}}$ to $\mathscr{S}$. (b) A specific choice of gauge may be imposed,
such that the combination of the constraints $C_{I}$ together with
the gauge-fixing conditions becomes second-class \cite{bojowald_canonical_2011}. One can thus obtain
a symplectic structure on the subspace of the constraint surface $\mathscr{S}\subset\mathscr{C}$
where the (appropriately chosen) gauge-fixing conditions are satisfied,
and where one will thus have a symplectic form $\bm{\omega}|_{\mathscr{S}}$.

In other to keep our discussion general, we use the notation $(\mathscr{S},\bm{\omega}|_{\mathscr{S}})$
to refer to the ``reduced phase space'' irrespective of whether
procedure (a) or (b) is used to define it.


\bibliography{references_v2}

\end{document}